\documentclass[preprint]{emulateapj}
\usepackage{apjfonts}
\usepackage{graphicx}

\newcommand{\kms}{km s$^{-1}$}

\received{}
\revised{}
\accepted{}

\slugcomment{}
\shorttitle{Late-time Light Curves of Type {\sc ii} SNe}
\shortauthors{Otsuka et al.}

\begin{document}

\title{Late-time Light Curves of Type {\sc ii} Supernovae: Physical
Properties of SNe and their environment}

\author{
Masaaki Otsuka\altaffilmark{1,2},
Margaret Meixner\altaffilmark{1},
Nino Panagia\altaffilmark{1},
Joanna Fabbri\altaffilmark{3},
Michael J. Barlow\altaffilmark{3},
Geoffrey C. Clayton\altaffilmark{4},
Joseph S. Gallagher\altaffilmark{5},
Ben E. K. Sugerman\altaffilmark{6},
Roger Wesson\altaffilmark{3},
Jennifer E. Andrews\altaffilmark{4},
Barbara Ercolano\altaffilmark{7}, and
Douglas Welch\altaffilmark{8}
}
\altaffiltext{1}{Space Telescope Science Institute, 3700 San Martin Dr.,Baltimore, MD 21218, USA; e-mail: {\tt otsuka@stsci.edu}}
\altaffiltext{2}{(current address) Institute of Astronomy and Astrophysics, Academia Sinica
P.O. Box 23-141, Taipei 10617, Taiwan, R.O.C.;  e-mail: {\tt otsuka@asiaa.sinica.edu.tw}}
\altaffiltext{3}{Department of Physics and Astronomy, University College London, Gower Street, London WC1E 6BT, UK}
\altaffiltext{4}{Department of Physics and Astronomy, Louisiana State
University, Baton Rouge, LA 70803}
\altaffiltext{5}{Department of Mathematics, Physics, and Computer Science, Raymond Walters College, 9555 Plain field
Rd., Blue Ash, OH 45236}
\altaffiltext{6}{Department of Physics and Astronomy, Goucher College, 1021 Dulaney Valley Road, Baltimore, MD 21204}
\altaffiltext{7}{Universit\"ats-Sternwarte M\"unchen, Scheinerstr. 1, 81679 M\"unchen,
Germany}
\altaffiltext{8}{Department of Physics and Astronomy, McMaster University, Hamilton, Ontario L8S 4M1, Canada}

%=======================================================================
\begin{abstract}
%=======================================================================
We present $BVRIJHK$ band photometry of 6 core-collapse supernovae,
SNe 1999bw, 2002hh, 2003gd, 2004et, 2005cs, and 2006bc measured at late epochs ($>$2 yrs)
based on Hubble Space Telescope ($HST$), Gemini north, and WIYN telescopes. We also show 
the $JHK$ lightcurves of a supernova impostor SN 2008S up to day 575 because 
it was serendipitously in our SN 2002hh field of view.  Of our 43 $HST$
observations in total, 36 observations are successful in detecting the light from
the SNe alone and measuring magnitudes of all the targets.  $HST$ observations
show a resolved scattered light echo around SN 2003gd at day 1520 and around SN 2002hh at day 1717. 
Our Gemini and WIYN observations detected SNe 2002hh and 2004et, as well.
Combining our data with previously published data, we show $VRIJHK$-band
 lightcurves and estimate decline magnitude rates at each band in 4 different phases.  Our prior work 
on these lightcurves and other data indicate that dust is forming in our targets from day $\sim$300-400, supporting
SN dust formation theory.  In this paper we focus on other physical
 properties derived from the late time light curves.  
We estimate $^{56}$Ni masses for our targets (0.5-14 $\times$10$^{-2}$ $M_{\odot}$) from the bolometric 
lightcurve of each for days $\sim$150-300 using SN 1987A as a standard (7.5$\times$10$^{-2}$ $M_{\odot}$). 
The flattening or sometimes increasing fluxes in the late time light curves of SNe 2002hh, 2003gd, 2004et 
and 2006bc indicate the presence of light echos.  We estimate the circumstellar hydrogen density of the material causing the light echo and find that SN 2002hh is surrounded by
 relatively dense materials ($n$(H) $>$400 cm$^{-3}$) 
and SNe 2003gd and 2004et have densities more typical of the interstellar medium ($\sim$1 cm$^{-3}$).  
We analyze the sample as a whole in the context of physical properties derived in prior work. 
The  $^{56}$Ni mass appears well correlated with progenitor mass with a slope of 0.31$\times$10$^{-2}$, supporting 
the previous work by Maeda et al. (2010), who focus on more massive Type {\sc ii} SNe. 
The dust mass does not appear to be correlated with progenitor mass.

%=======================================================================
\end{abstract}
%=======================================================================

\section{Introduction}
Since the discovery of SN 1987A, dust formation has been
confirmed observationally in the ejecta of several core-collapse
supernovae (Type {\sc ii} SNe; e.g., Wooden et al. 1993, Kozasa et at
al. 2009).
In these, one or more of the following signatures are observed
$\sim$250-600 days after the initial explosion: (1) the development
of an infrared excess seen in the spectral energy distribution
(SED) of the SN; (2) the appearance of blue-shifted
optical/near-infrared lines, which is interpreted as the attenuation
of the redshifted wing (i.e., the receding gas) by dust formed
homogeneously in the ejecta; and
(3) the steepening decline of the optical lightcurve consistent with an
increase in circumstellar extinction.
The SN luminosity after day $\sim$150 is supported by $\gamma$ rays from $^{56}$Co decaying into  $^{56}$Ni.
If the dust is formed in SNe, the optical luminosity from the SN remnant
has become much fainter than would be produced by the radioactive decay power.

In order to properly assess dust formation by Type {\sc ii} SNe ejecta it is essential to continue monitoring
Type {\sc ii} SNe before and after dust formation started (i.e. day $\sim$300) because other physical properties such 
as the $^{56}$Ni mass and light echos from circumstellar and
interstellar dust can be measured and disentangled from dust production effects. 
The early epoch evolution of Type {\sc ii} SNe before day $\sim$300 has been relatively well studied in optical and near-IR bands.
However at the present time, the evolution beyond day 300 has been poorly measured.

To address these issues, we have conducted a long term photometric monitoring of SNe 
using high angular resolution facilities such as HST/WFPC2/NICMOS, WIYN/WHIRC, 
and Gemini/NIRI through our SNe project $SEEDS$ (Survey for Evolution of Emission from Dust in SNe, PI: M.Barlow). 
In this paper we report on the late-epoch light-curves of 6 Type {\sc ii} SNe. 
The remainder of this paper is organized as follows.  In section 2, we present 
the observations, data processing and resulting images, and 
light curves. In section 3,  we analyze the resulting light curves to derive $^{56}$Ni masses for all targets except for 
SN 1999bw and, for cases with detected light echoes, to estimate the density of the environment. 
In section 4, we discuss the results for each SN.  In section 5, we do a comparative analysis of these SNe 
in the context of physical properties reported in the literature, such
as the amount of dust produced and progenitor mass.

\section{Observations \& Data reduction}
\subsection{Sample selection}
We chose 6 Type {\sc ii} SNe in nearby galaxies, matching the following three criteria.
(1) The distance to a SN is within $\sim$15 Mpc and thus suitable for long-term follow-up
observations with $HST$, $Spitzer$, and $>$4-m class ground based
telescopes.  (2) At the start of the observations, the SN is 1 year
or older after the initial explosion. (3) The expected $V$-band magnitude of targets is $<$28  based on the evolution of SN 1987A in order
that the SNe is detectable within a reasonable integration time.
These 6 Type {\sc ii} SNe include 4 Type {\sc ii}-P (SNe 2002hh,
2003gd, 2004et, and 2005cs), 1 Type {\sc ii}-n (SN 1999bw), and 1 
Type {\sc ii}-L or {\sc ii}-P (SN 2006bc).  In addition, during our
campaign, a supernova impostor, SN 2008S, erupted in the field of view of our SN 2002hh observations, allowing us to monitor this new type of variable as well.

\subsection{$HST$ observations}
The high-angular resolution images of our target SNe were taken by the
Hubble Space Telescope with the Wide Field Planetary Camera 2
(WFPC2) and the Near Infrared Camera and Multi-Object Spectrometer 2
(NICMOS2) through the cycle 16 proposal by M.Meixner (proposal
ID: GO 11229). The observations were coordinated so that the WFPC2 and NICMOS
images were taken within days to a week of each other.
In the WFPC2 observations, we took the images using the 
F450W, F606W, F622W, and F814W filters. In the NICMOS2 observations,
we took the images using the F110W, F160W, and F205W filters.
The observation logs and SN positions are summarized in Table \ref{obslog}. 
The boldface coordinates in this table are 
derived from our F606W or F622W $HST$ observations and the others 
are from the SIMBAD database.

\subsection{$HST$ data reduction}

The data reduction and calibration were performed by standard 
techniques. Using the STSDAS package version 3.8 specialized for 
$HST$, we reduced and calibrated the data, including cosmic rays and
artificial features removal and detector linearity correction. The final
high resolution images were created 
using the STSDAS/DRIZZLE package. For both the WFPC2 and
NICMOS2 observations, we conducted sub pixel scale dithering to improve
S/N and remove cosmic rays and obtain further high-resolution images via the
drizzle technique. We set 4 or 5 points dithering with 115-400sec per 
science frame for each band.  After
DRIZZLE processing, we performed an additional distortion
correction and alignment using background stars for the NICMOS2 data.
The point-spread-function (PSF) of these reference background
stars were fit by Gaussian profiles to obtain accurate positions,
and then instrumental distortions were corrected with the XYXYMATCH,
GEOMAP, and GEOTRAN IRAF routines, and we aligned the positions of stars in the NICMOS
detector to those in the WFPC2. The resultant plate scale 
reaches 0.025$''$ pixel$^{-1}$ in the WFPC2 and archived ACS/HRC data (see below) and
0.04$''$ pixel$^{-1}$ in the NICMOS2 and archival NICMOS3 data (see below), respectively.
The size of the PSF at FWHM typically corresponds to $\sim$3
pixel. Using the PSF fitting package IRAF/DAOPHOT, photometry was 
performed on the images, except for  SN 2005cs in F110W and F160W 
and SN 2002hh in all bands. For SN 2005cs in F110W and F160W, we used differential 
images with pre-explosion images using archival NICMOS3 images of the host galaxy taken 
in 1998 June 28th (PI:N.Scoville, Proposal-ID: GO 7237) and 
our NICMOS2 data 
to detect the SN alone and performed PSF fitted photometry.   SN 2002hh has an extended 
shell structure as shown in Fig.1,  we therefore performed aperture photometry based 
on background star subtracted images. We used 0.6$''$ radius regions to define the entire 
the SN. We subtracted background from an annulus centered on the SN with inner and outer 
radii of 0.6$''$ and 0.8$''$.  The aperture corrections were calculated
with an empirical PSF function. Note that the magnitudes of SN2002hh
listed in Tables 1, 2, and 3 include the SN + extend light echo.

For each filter, the measured count rate (CR, in units of DN s$^{-1}$)
of the SN was converted to flux by multiplication with the PHOTFLAM (erg
s$^{-1}$ cm$^{-2}$ {\AA}$^{-1}$ DN$^{-1}$ or erg
s$^{-1}$ cm$^{-2}$ {\AA}$^{-1}$ electron$^{-1}$) conversion factor given in
the fits header, where PHOTFLAM is the bandpass-averaged flux density
for a source that would produce a count rate (CR in units of DN s$^{-1}$ or electron s$^{-1}$).
The $HST$ system magnitude (STMAG) was
calculated from the following equation,
\begin{equation}
STMAG = -2.5\log_{10}(CR \times PHOTFLAM) + ZP,
\end{equation}
\noindent where ZP is the zero point magnitude for 1 DN s$^{-1}$ or 1 electron s$^{-1}$.
For the WFPC2 and archival ACS data (see below), the conversion from F450W, F606W and F622W, and F814W STMAG
to the Vega magnitude system $B$, $V$, $R$, and $I$, respectively, used the STSDAS/SYNPHOT package.
The broad band NICMOS filters, F110W, F160W and F205W, are
roughly equivalent to the $J$, $H$ and $K$ filters respectively, color
transformations are
not well constrained for late-time SN spectra at near-IR wavelengths.
Based on our experience, for SNe in days 300-1000, the spectral energy distribution (SED)
in the range from $\sim$0.44 to $\sim$2.2 $\mu$m can be
represented by a single blackbody function with the effective temperature
of $\sim$7\,000-11\,000 K. The conversion from the STMAG to the
Vega magnitude system was done by the STSDAS/SYNPHOT package, assuming
a 9\,000 K blackbody function as the incident SED.

\begin{table*}[t]
\scriptsize
\centering
\caption{$HST$ Observation log and measured magnitudes}
\begin{tabular}{@{}l@{\hspace{6pt}}c@{\hspace{6pt}}c@{\hspace{6pt}}c@{\hspace{6pt}}
c@{\hspace{6pt}}c@{\hspace{6pt}}c@{\hspace{6pt}}l@{\hspace{6pt}}c@{\hspace{6pt}}l@{\hspace{6pt}}c@{\hspace{6pt}}c@{\hspace{6pt}}c@{}}
\hline\hline
SN Name &Host Galaxy &$D$(Mpc)&$E(B-V)$&$\alpha$(2000.0) & $\delta$(2000.0) & Obs.Date & Epoch (day) &
 Exp.time  & Detector & Filter&Detect&Magnitudes\\
%(1)&(2)&(3)&(4)&(5)&(6)&(7)&(8)&(9)&(10)\\
\hline
SN1999bw&NGC 3198&13.7&0.013 &10:19:46.81  & +45:31:35.0   & 2008/04/03   & 3271.24 & 128sec$\times$5   & NIC2  & F110W&N&\nodata   \\
({\sc ii}-n) &  &&&{\bf 10:19:46.68}  &{\bf +45:31:37.32} & 2008/04/03   & 3271.25 & 128sec$\times$4   & NIC2  & F160W&N&\nodata    \\
 &  &  &&& &2008/04/03   & 3271.26 & 115sec$\times$5   & NIC2  & F205W&N&\nodata    \\
 &  &  &&& &2008/04/03   & 3271.11 & 400sec$\times$4   & PC1  & F606W&Y&{\bf 25.77 $\pm$ 0.16} \\
 &  &  &&& &2008/04/03  & 3271.20 & 400sec$\times$4   & PC1  & F814W&Y&{\bf 25.55 $\pm$ 0.24}  \\
SN2002hh&NGC 6946&5.9&1.97&20:34:44.29             &+60:07:19.0 & 2007/07/09   & 1716.88 & 400sec$\times$4   & PC1  & F606W&Y&{\bf 20.62 $\pm$ 0.11}  \\
({\sc ii}-P)          &&&  &{\bf 20:34:44.19} &{\bf +60:07:20.11}   & 2007/07/09  & 1716.91 & 400sec$\times$4   & PC1  & F814W&Y&{\bf 19.67 $\pm$ 0.11}  \\
 &  &  &&&& 2007/07/09 & 1716.96 & 400sec$\times$4   & PC1  & F450W&Y&{\bf 23.25 $\pm$ 0.18} \\
 &  &  &&&& 2007/07/10   & 1717.02 & 128sec$\times$5   & NIC2  & F110W&Y&{\bf 18.84 $\pm$ 0.11}  \\
 &  &  &&&& 2007/07/10   & 1717.03 & 128sec$\times$4   & NIC2  & F160W&Y&{\bf 18.53 $\pm$ 0.11}  \\
 &  &  &&&& 2007/07/10   & 1717.04 & 115sec$\times$5   & NIC2  & F205W&Y&{\bf 17.44 $\pm$ 0.11}  \\
SN2003gd  &NGC 628 &7.2&0.18&01:36:42.65  & +15:44:19.9  & 2007/08/11& 1519.79 &400sec$\times$4   & PC1  & F622W&Y&{\bf 24.91 $\pm$ 0.18}  \\
({\sc ii}-P) &&  &&  {\bf 01:36:42.71}&{\bf +15:44:20.25}& 2007/08/11  & 1519.86 & 400sec$\times$4   & PC1  & F814W&Y&{\bf 25.60 $\pm$ 0.19}   \\
 &  &  &&& &2007/08/14  & 1523.25 & 128sec$\times$5   & NIC2  & F110W&Y&{\bf $<$24.92}  \\
 &  &  &&& &2007/08/14  & 1523.26 & 128sec$\times$4   & NIC2  & F160W&Y&{\bf $<$25.47}  \\
 &  &  &&& &2007/08/14  & 1523.27 & 115sec$\times$5   & NIC2  & F205W&N&\nodata   \\
SN2004et  &NGC 6946&5.9&0.41 &20:35:25.33  & +60:07:17.7  & 2007/07/08  & 1071.74 & 400sec$\times$4   & PC1  & F606W&Y&{\bf 23.01 $\pm$ 0.12}  \\
({\sc ii}-P) &  &&&  {\bf 20:35:25.36}& {\bf +60:07:17.80} & 2007/07/08  & 1071.82 & 400sec$\times$4   & PC1  & F814W&Y&{\bf 23.04 $\pm$ 0.14}  \\
 &  &  &&& &2007/07/08  & 1071.91 & 128sec$\times$5   & NIC2  & F110W &Y&{\bf 21.76 $\pm$ 0.14}   \\
 &  &  &&& &2007/07/08  & 1071.94 & 128sec$\times$4   & NIC2  & F160W &Y&{\bf 22.56 $\pm$ 0.22}  \\
 &  &  &&& &2007/07/08  & 1071.95 & 115sec$\times$5   & NIC2  & F205W&Y&{\bf $<$22.07}  \\
 &  &  &&& &2008/01/19  & 1266.71 & 400sec$\times$4   & PC1  & F606W &Y&{\bf 23.35 $\pm$ 0.11} \\
 &  &  &&& &2008/01/19  & 1266.78 & 400sec$\times$4   & PC1  & F814W&Y&{\bf 23.35 $\pm$ 0.13}  \\
 &  &  &&& &2008/01/20  & 1267.58 & 128sec$\times$5   & NIC2  & F110W&Y&{\bf 21.99 $\pm$ 0.15}   \\
 &  &  &&& &2008/01/20   & 1267.59 & 128sec$\times$4   & NIC2  & F160W &Y&{\bf 22.62 $\pm$ 0.24}  \\
 &  &  &&& &2008/01/20  & 1267.59 & 115sec$\times$5   & NIC2  & F205W&Y&{\bf 21.07 $\pm$ 0.19}    \\
SN2005cs  &NGC 5194&7.1&0.14 &13:29:53.37  & +47:10:28.2  & 2007/10/28  & 851.96 & 400sec$\times$4   & PC1  & F606W&Y&{\bf 25.16 $\pm$ 0.19}  \\
({\sc ii}-P) &&  &&  {\bf 13:29:52.80}&{\bf +47:10:35.64} & 2007/10/28  & 852.02 & 400sec$\times$4   & PC1  & F814W &Y&{\bf 25.07 $\pm$ 0.24}   \\
 &  &  &&& &2007/10/31  & 854.15 & 128sec$\times$5   & NIC2  & F110W&Y&{\bf $<$23.30} \\
 &  &  &&& &2007/10/31  & 854.16 & 128sec$\times$4   & NIC2  & F160W&N&\nodata   \\
 &  &  &&& &2007/10/31  & 854.17 & 115sec$\times$5   & NIC2  & F205W&N&\nodata  \\
 &  &  &&& &2008/03/31  & 1006.15 & 400sec$\times$4   & PC1  & F606W&Y&{\bf $<$25.94}  \\
 &  &  &&& &2008/03/31  & 1006.21 & 400sec$\times$4   & PC1  & F814W&N&\nodata  \\
SN2006bc  &NGC 2397&16&0.52& 07:21:16.50  & --68:59:57.3  & 2007/09/07  & 531.98 & 400sec$\times$4   & PC1  & F606W&Y&{\bf 23.16 $\pm$ 0.19}  \\
({\sc ii}-P or L)&  &&  &{\bf 07:21:16.45}&{\bf --68:59:56.75} & 2007/09/07  & 532.07 & 400sec$\times$4   & PC1  & F814W&Y&{\bf 21.72 $\pm$ 0.19} \\
 &  &&  &&& 2007/09/17  & 541.45 & 128sec$\times$5   & NIC2  & F110W&Y&{\bf 21.60 $\pm$ 0.20}  \\
 &  &&  &&& 2007/09/17   & 541.46 & 128sec$\times$4   & NIC2  & F160W&Y&{\bf 20.80 $\pm$ 0.20}  \\
 &  &&  &&& 2007/09/17   & 541.47 & 115sec$\times$5   & NIC2  & F205W&Y&{\bf 19.90 $\pm$ 0.10} \\
 &  &&  &&& 2008/02/17  & 694.25 & 400sec$\times$4   & PC1  & F606W&Y&{\bf 23.85 $\pm$ 0.26}  \\
 &  &&  &&& 2008/02/17  & 694.28 & 400sec$\times$4   & PC1  & F814W&Y&{\bf 22.51 $\pm$ 0.31}   \\
 &  &&  &&& 2008/02/17  & 694.35 & 128sec$\times$5   & NIC2  & F110W&Y&{\bf 22.40 $\pm$ 0.30}  \\
 &  &&  &&& 2008/02/17  & 694.36 & 128sec$\times$4   & NIC2  & F160W&Y&{\bf 21.70 $\pm$ 0.40}   \\
 &  &&  &&& 2008/02/17  & 694.37 & 115sec$\times$5   &NIC2  & F205W&Y&{\bf 20.70 $\pm$ 0.20}  \\
\hline
\end{tabular}
\tablerefs{(For $D$ and $E(B-V)$) SN 1999bw: Freedman et al. (2001), Smith et al. (2010); 
SN 2002hh: Barlow et al. (2005), Karaehentsev et al. (2000), Pozzo et al. (2006); SN 2003gd: 
Van Dyk et al. (2003), Hendry et al. (2005); SN 2004et: Fabbri et al. (2011a), Karaehentsev 
et al. (2000); SN 2005cs: Tak{\'a}ts, K., \& Vink{\'o} (2006), Tsvetkov et al. (2006); 
SN 2006bc: Gallagher et al. (2010); Mihos \& Bothun (1997)}
\tablecomments{The measured magnitudes at F450W, F606W, F622W, F814W, F110W, F160W,
and F205W in column (10) are converted to $B$, $V$, $R$, $I$, $J$, $H$, $K$ magnitudes. 
 We converted the $STMAG$ into the Johnson-Cousins system and Vega system using the STSDAS/SYNPHOT. 
In the fifth and sixth columns: the boldface characters are the SN positions derived from our $HST$ observations and the other characters 
are from SIMBAD database. In the tenth column, PC1 and NIC2 mean the PC1 chip in the WFPC2 and 
the NICMOS2 chip in the NICMOS, respectively (see text for detail).\label{obslog}}
\end{table*}

\begin{figure*}[t]
\centering
\includegraphics[scale=1.3]{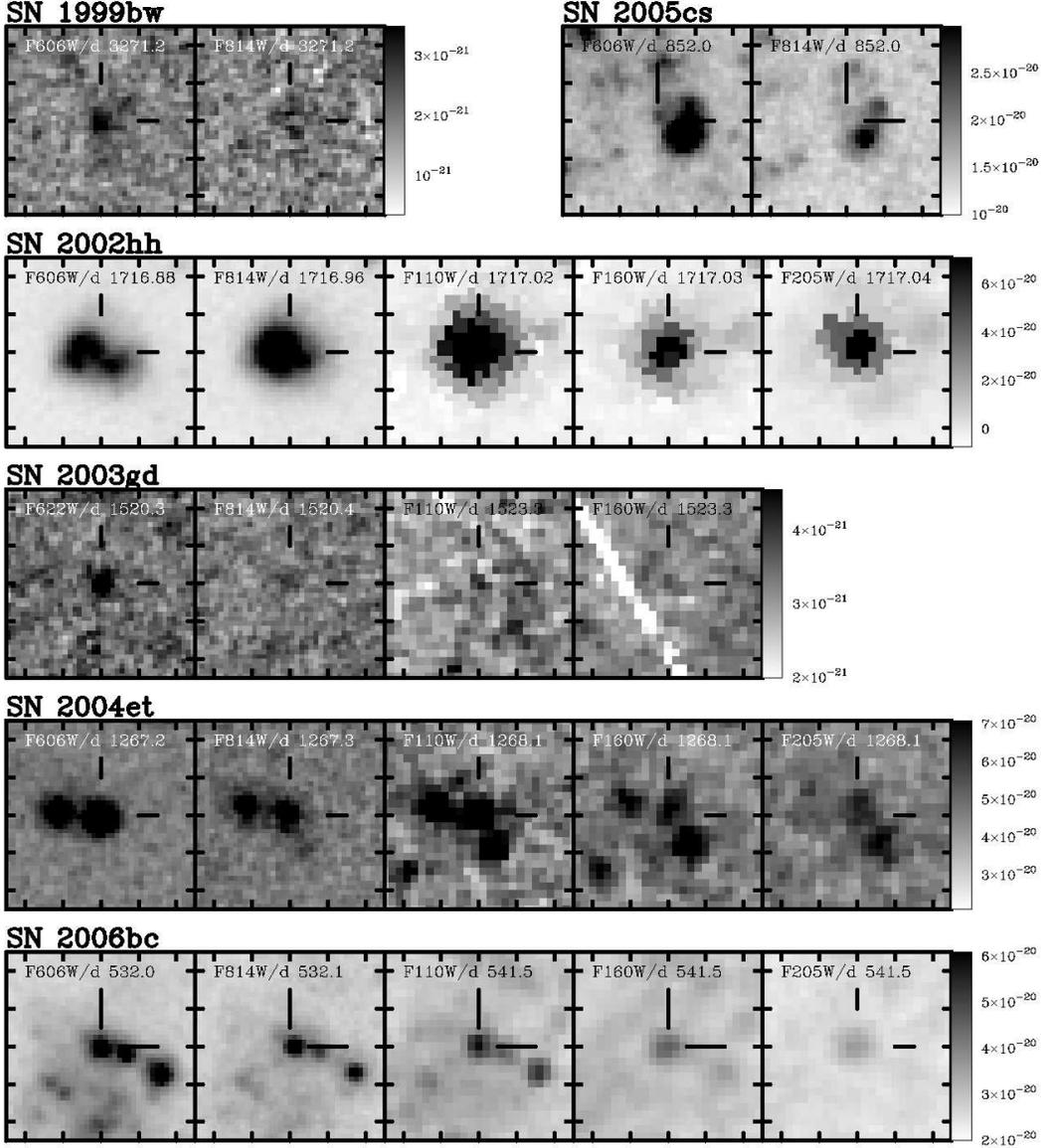}
\caption{The {\it HST} images of SNe 1999bw, 2002hh, 2003gd, 2004et,
 2005cs, and 2006bc. The SN is in the center of each panel.
The size of each panel is 1\arcsec $\times$ 1\arcsec and each tick mark is  0.2\arcsec. 
The intensity scale (erg s$^{-1}$ cm$^{-2}$ {\AA}$^{-1}$ pixel$^{-1}$) is common in each SN. In all cases north is up and east is to the left. Each box has a label for HST filter and  day after explosion.}
\label{hstim}
\end{figure*}

\subsection{$HST$ Archive data}
Except for SN 2004et,  archival WFPC2, Advanced Camera for Surveys (ACS), and NICMOS3 images are available for our SNe sample as noted in Table 2.
We created drizzled images and
we measured the magnitudes by the same way as described above. The measured magnitudes are listed in Table \ref{prehst}.
In this table we also list the magnitudes measured  at earlier epochs as found in the literature as noted in the table.

\begin{table}[h]
\centering
\scriptsize
\caption{The magnitudes by previous $HST$ observations.\label{prehst}}
\begin{tabular}{@{}lclcccl@{}}
\hline\hline
{SN Name} &
{Obs.Date} &
{Epoch}  &
{Detector}&
{Filter} &
{Magnitudes} &
{Ref.} \\
\hline
SN 1999bw&2001/01/09&629.88      &PC1       &F555W($V$)   &24.08 $\pm$ 0.06 &(1)\\
SN 2002hh&2005/09/17&1056.26     &HRC       &F606W($V$)   &20.88 $\pm$ 0.11 &(2)\\
        &2005/09/17&1056.40     &HRC       &F814W($I$)  &19.49 $\pm$ 0.11 &(2)\\
        &2006/04/23&1273.94     &HRC       &F606W($V$)   &20.91 $\pm$ 0.11 &(2)\\
SN 2003gd&2004/12/08&544.11      &HRC       &F435W($B$)   &23.98 $\pm$ 0.15 &(3)\\
        &2004/12/08&544.11      &HRC       &F622W($R$)   &23.13 $\pm$ 0.16 &(3)\\
        &2007/06/21&1469.45     &PC1       &F450W($B$)   &25.01 $\pm$ 0.17 &(4)\\
        &2007/06/21&1469.45     &PC1       &F675W($R$)   &24.10 $\pm$ 0.16 &(4)\\
SN 2005cs&2006/12/10&529.02      &HRC       &F555W($V$)   &22.07 $\pm$ 0.15 &(5)\\
        &2006/12/10&529.02      &HRC       &F814W($I$)  &21.36 $\pm$ 0.15 &(5)\\
SN 2006bc&2006/10/14&203.28      &WFC1       &F555W($V$)  &22.10 $\pm$ 0.10 &(6)\\
        &2006/10/14&203.28      &WFC1       &F814W($I$)  &19.80 $\pm$ 0.10 &(6)\\
\hline
\end{tabular}
\tablecomments{HRC means High Resolution Camera mode in the ACS. WFC1 means the wide field camera 1 chip in the WFPC2.}
\tablerefs{
(1) Li et al. (2002); (2) This work (data source; GO 10607, PI: Sugerman); (3) Sugerman (2005);
(4) Maund \& Smartt (2009); (5) This work (data source; GO 10877, PI: Li); (6) Gallagher et al. (2011; GO 14980, PI: Smartt)
}
\end{table}

\begin{table}
\centering
\scriptsize
\caption{NIRI and WHIRC observations.\label{niri}}
\begin{tabular}{@{}crcccc@{}}
\hline\hline
\multicolumn{6}{c}{SN 2002hh}\\
Obs.Date&Epoch&Instr.&$J$&$H$&$K$\\
\hline
2005/06/04&951.5   &NIRI&18.50 $\pm$ 0.16&17.76 $\pm$ 0.16&16.93 $\pm$ 0.20\\
2005/08/04&1012.5  &NIRI&18.57 $\pm$ 0.15 &17.80 $\pm$ 0.16 &16.93 $\pm$ 0.20\\
2005/10/16&1085.3  &NIRI&$>$18.48&$>$17.88&16.78 $\pm$ 0.20\\
2006/07/07&1349.5  &NIRI&18.51 $\pm$ 0.15 &17.75 $\pm$ 0.16 &16.95 $\pm$ 0.20 \\
2007/07/10&1717.0&HST&18.61 $\pm$ 0.19 &18.03 $\pm$ 0.23&17.02 $\pm$ 0.21\\
2008/11/12&2208.1  &WHIRC&18.59 $\pm$ 0.15&17.82 $\pm$ 0.16&17.00 $\pm$ 0.20\\
2009/05/10&2387.4  &WHIRC&\nodata    &17.54 $\pm$ 0.20&\nodata\\
2009/08/30&2499.3  &WHIRC&\nodata    &18.07 $\pm$ 0.20&\nodata\\
\hline
\multicolumn{6}{c}{SN 2004et}\\
2009/08/30&1801  &WHIRC&\nodata    &$<$22.6&\nodata\\
\hline
\multicolumn{6}{c}{SN 2008S}\\
2008/11/12&283.3  &WHIRC&18.37 $\pm$ 0.15&16.73 $\pm$ 0.15&15.64 $\pm$ 0.20\\
2009/05/10&463.5  &WHIRC&\nodata    &17.83 $\pm$ 0.16&16.25 $\pm$ 0.73\\
2009/08/30&575.5  &WHIRC&\nodata    &18.74 $\pm$ 0.16&\nodata\\
\hline
\end{tabular}
\tablecomments{The errors of 2MASS stars' magnitudes are included. In SN 2002hh,
the $HST$ magnitudes were re-calibrated using the NIRI data and nearby 2MASS stars' magnitude.}
\end{table}

\subsection{NIRI and WHIRC observations}
To complement $HST$ observations, we performed $JHK$-band high-resolution imaging observations
for SNe 2002hh, 2004et, and 2008S using the Gemini north 8.2-m telescope with the Near Infrared Imager
and spectrometer (NIRI) and the WIYN 3.5-m telescope
with the WIYN High Resolution Infrared Camera (WHIRC; Meixner et al. 2010).
The NIRI and Aug 2009 WHIRC observations were carried out through the NOAO open use programs of P.Is.
G.C.Clayton and M.Otsuka, respectively. The other WHIRC observations were carried out in
STScI guaranteed observation time (P.I.: M.Meixner). The plate scale of both instruments
is $\sim$0.1$''$ pixel$^{-1}$ and the seeing was 0.3$''$-0.8$''$ at $JHK$-bands.
The average PSF size of $\sim$0.5$''$ enabled us
to measure magnitudes of both SNe with relatively little contamination from nearby stars.

The data reductions using the IRAF/MSCRED package were carried out in a standard manner (dark subtraction,
flat-fielding, and local sky subtraction). For the WHIRC data,
we performed a detector linearity correction using the WHIRC task WPREP and distortion correction using the files
downloaded from the WIYN/WHIRC web page. The magnitude
measurements were performed using the IRAF/DAOPHOT. The conversion of the instrumental magnitudes into
the standard system was performed relative to Two Micron All-Sky Survey (2MASS) photometry of 4-10 stars in the field.

The measured magnitudes are listed in Table \ref{niri}. The errors include the
estimated errors of 2MASS stars' magnitudes. For SN 2002hh, we converted the Vega magnitudes derived from $HST$ data to 2MASS
magnitudes. Comparison between $HST$ and 2MASS magnitudes indicates that a star nearby SN 2002hh contributes $\sim$0.2-0.3 mag.

\subsection{Results}

\begin{figure}[t]
\centering
\includegraphics[scale=0.42,bb=20 182 582 582]{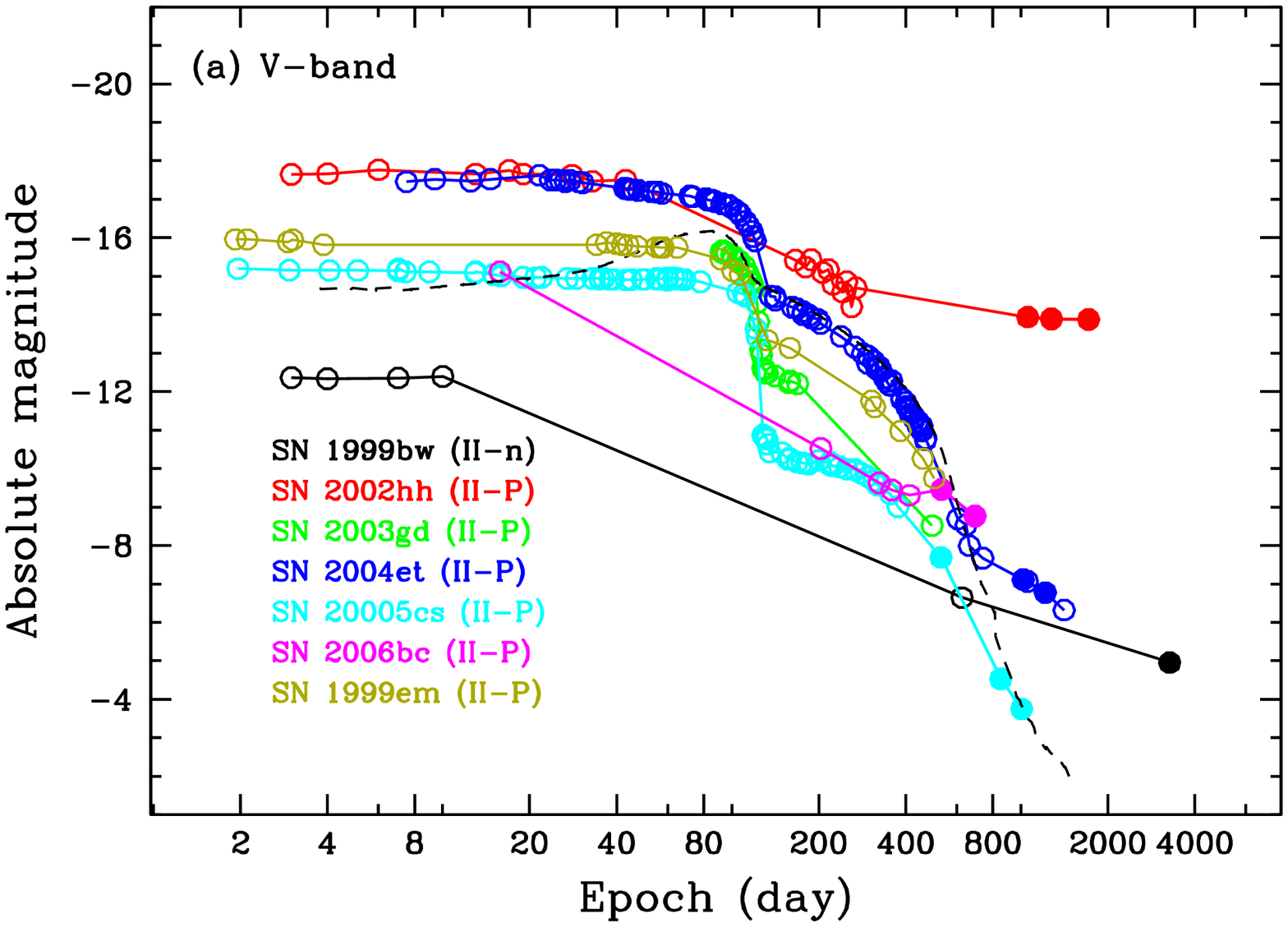}\\
\includegraphics[scale=0.42,bb=20 182 582 582]{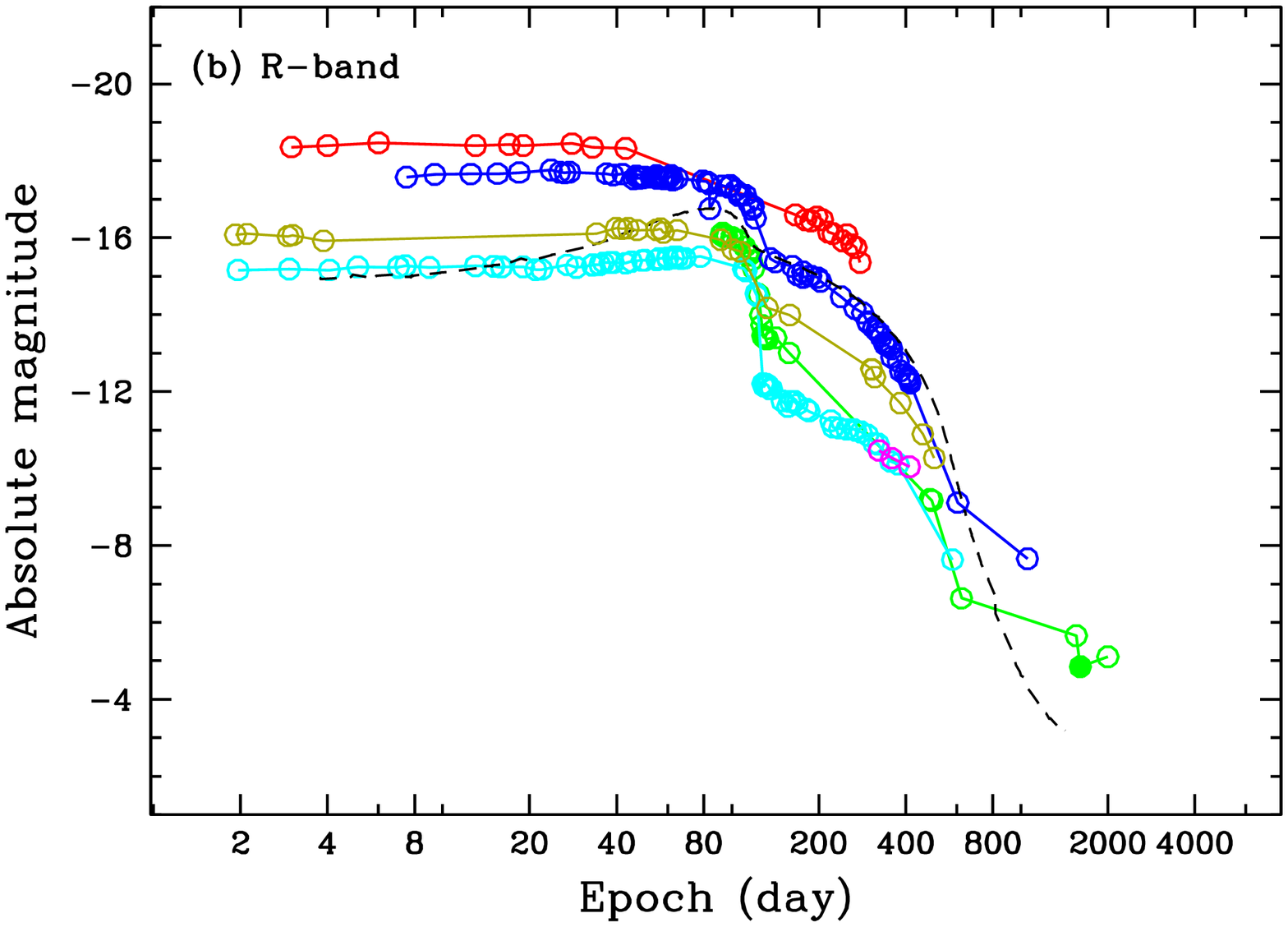}\\
\includegraphics[scale=0.42,bb=20 182 582 582]{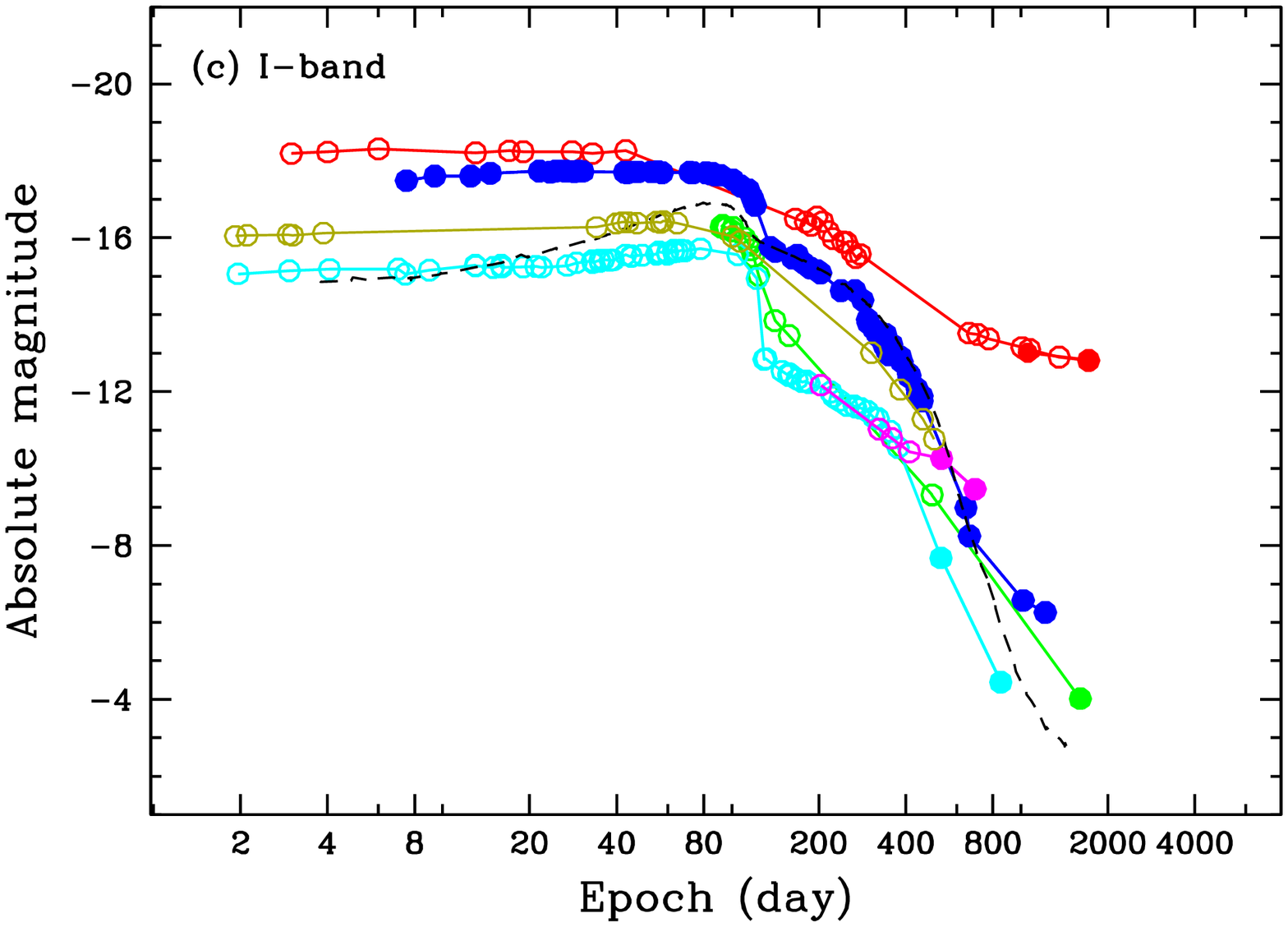}
\caption{$VRI$ lightcurves on a de-reddened absolute magnitude scale.
The open circles are from published papers and the filled circles indicate our contributions.
The broken lines are from published SN 1987A data. The data sources are listed in Table \ref{refer}. \label{lc}}
\end{figure}
\begin{figure}[t]
\centering
\includegraphics[scale=0.42,bb=20 182 582 582]{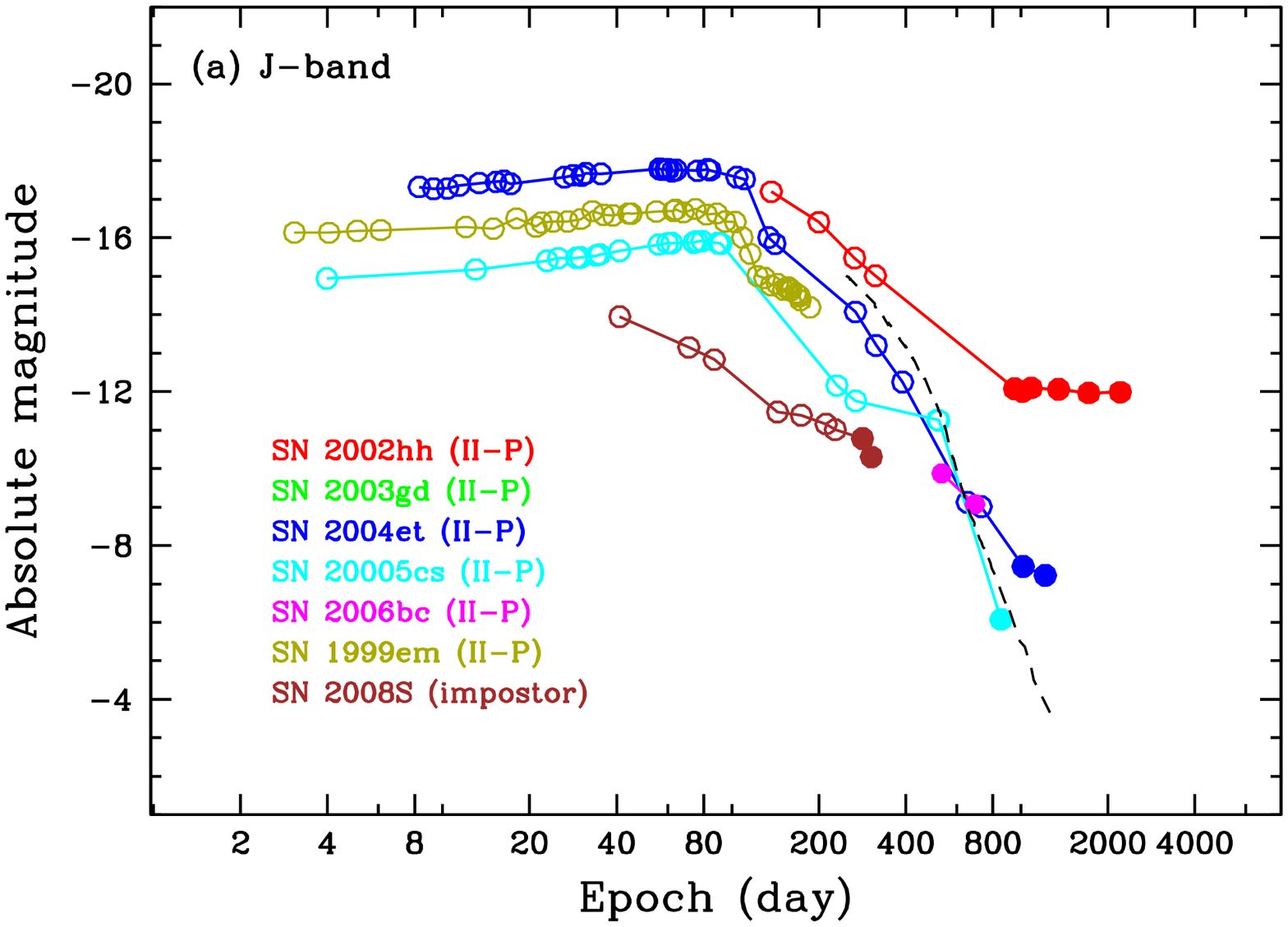}\\
\includegraphics[scale=0.42,bb=20 182 582 582]{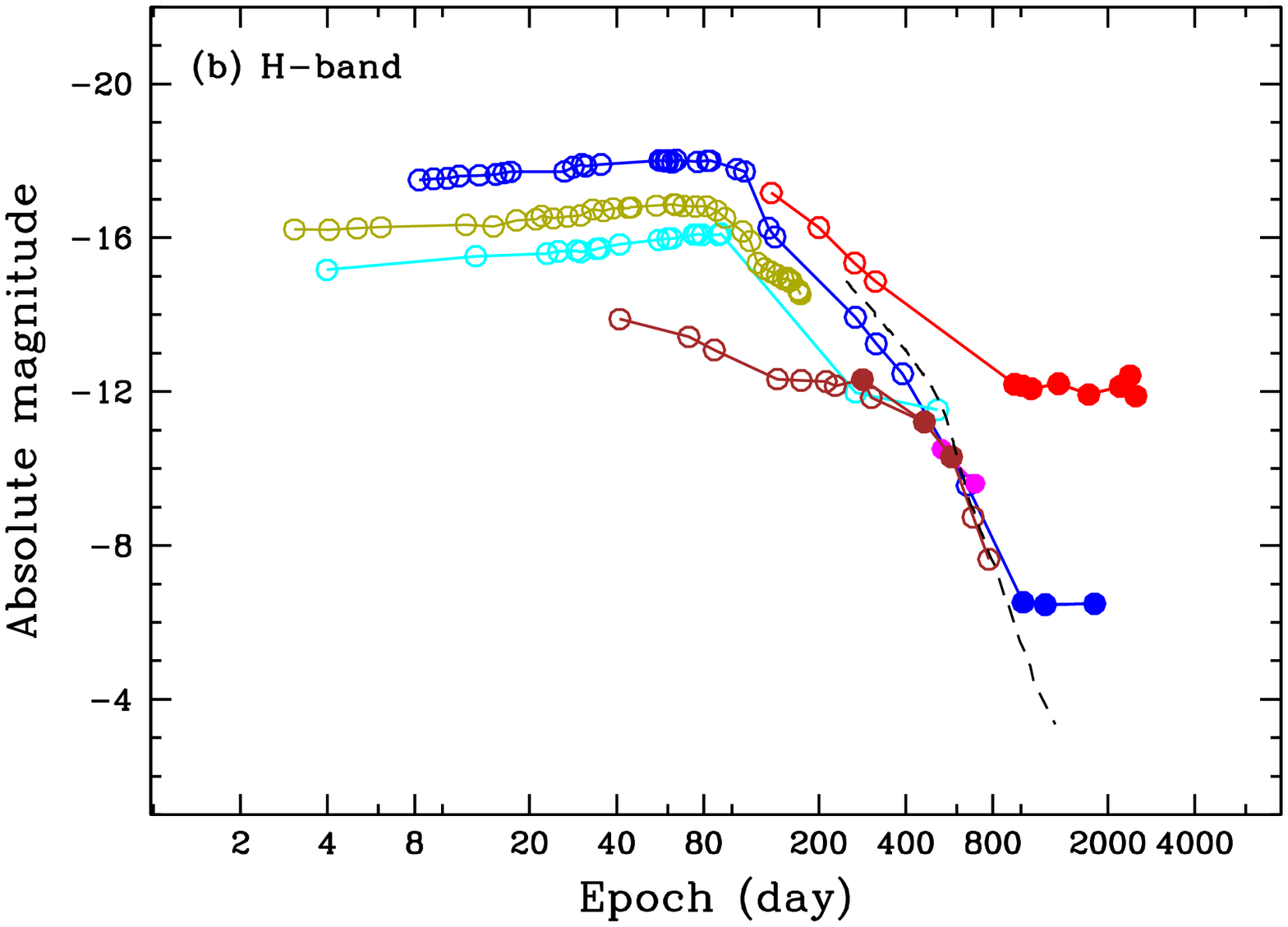}\\
\includegraphics[scale=0.42,bb=20 182 582 582]{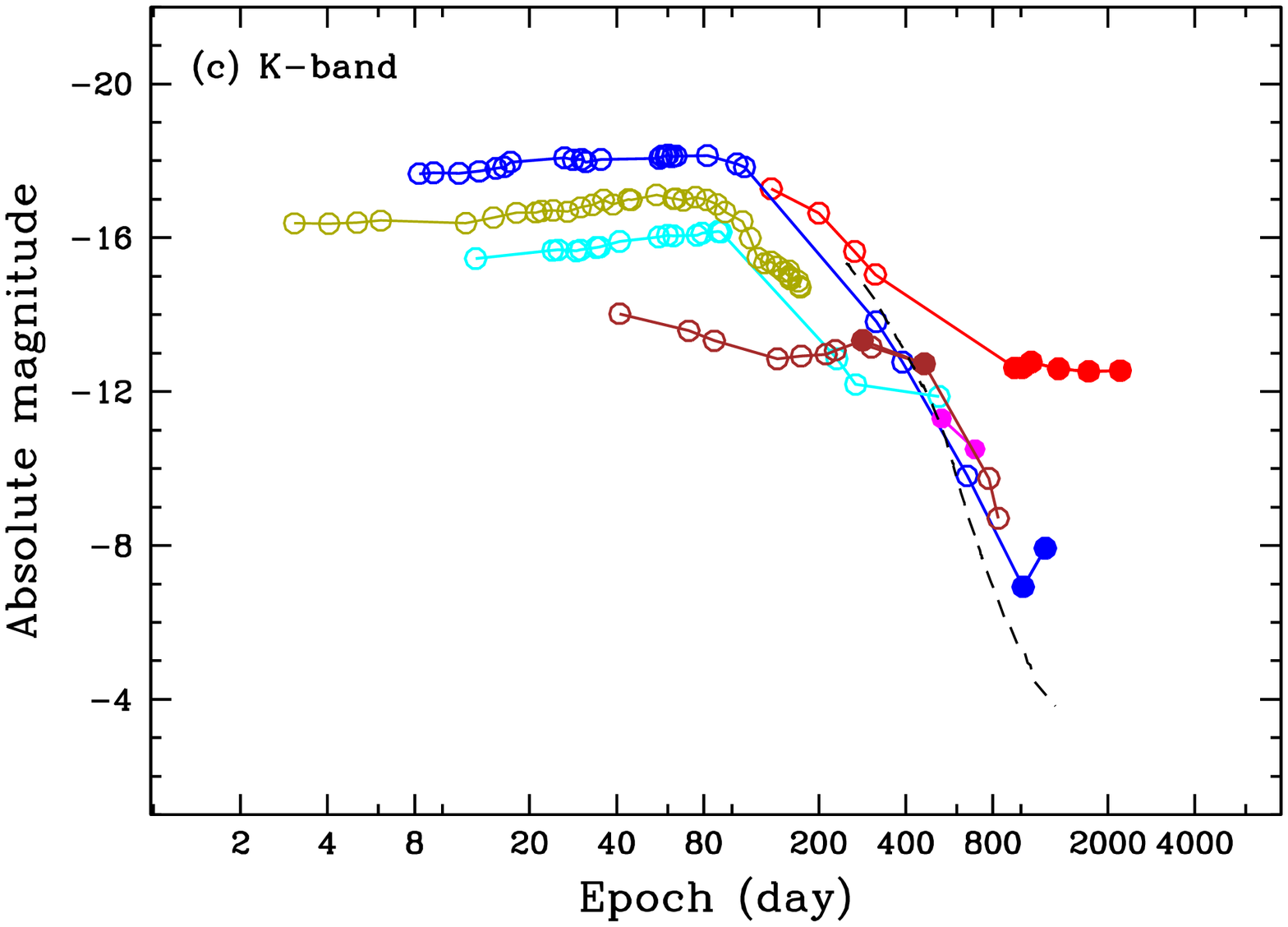}
\caption{The lightcurves at $JHK$-bands in de-reddened absolute magnitude scale. The symbols and lines are the same meanings in 
Fig.1. The data sources are listed in Table \ref{refer}. \label{lc2}}
\end{figure}

The resultant images of target SNe are presented in Fig.\ref{hstim}.
 We are successful in detecting the SNe  and
measuring either F606W($V$) or F622W($R$) band magnitudes of all the targets in $HST$ observations.
The light-curves of each SN are presented in Figs.\ref{lc} ($VRI$) and \ref{lc2} ($JHK$-bands).
Theses light-curves are based on our data (filled circles) and data from
 the literature for each object (open circles). The data after $>$600-1000 days are mainly 
from the former and the data before this phase are mainly from the latter. 
The data sources for each object are summarized in Table \ref{refer}. 
The values in Figs. \ref{lc} and \ref{lc2} are de-reddened absolute 
magnitudes adopting  the distance to the host galaxy, $D$, and $E(B-V)$ listed in the third and
forth columns of Table 1. 
All $E(B-V)$ in Table 1 are the total extinction by the Milky Way Galaxy and host galaxy.
We used the reddening function assuming the $R_{V}$=3.1 case
by Cardelli et al. (1989) for all SNe. As references, we plot the lightcurves of the
 Type {\sc ii}-P SNe 1987A (Hamuy \& Suntzeff 1990; 
Walker \& Suntzeff 1991) and 1999em (Elmhamdi et al. 2003), which are well studied Type {\sc ii} SNe. 
 
We measured the rate of fading of SNe $\gamma_{band}$ during 4 phases, namely,
days $\sim$30-100 (plateau phase), $\sim$150-300 (radioactive decay phase),
$\sim$300-800 (dust forming phase), and $>$800. The results are summarized 
in Table \ref{dlm}. For each phase, we fit a line to the light-curve, the slope of which is 
the fading rate for that phase measured in units of magnitude per 100 days. 
The values indicated by bold face are derived from the data presented in this work.
We note that light echoes, which may appear as flattening or increasing fluxes in the 
lightcurve, could arise during any of these phases depending on the location of 
the scattering material with respect to the SNe. The lightcurves of Type {\sc ii} SNe during the radioactive decay 
phase can be explained by $\gamma$-rays power by the radioactive decay of 
$^{56}$Co to $^{56}$Fe, at a rate corresponding to the $e$-folding life time of 
the $^{56}$Co decay ($t_{56}$ = 111.3 days). The rate expected for radioactive decay of $^{56}$Co into $^{56}$Fe is 0.98 mag per 100 days, while the averaged $\gamma_{V}$ among 8 SNe is 0.96$\pm$0.12 
(Patat et al. 1994).
For SN 2004et, for example, Sahu et al. (2006) found 
that the $BVRI$ lightcurves in 180-310 days were linear with  $\gamma_{V}$ = 1.04, 
$\gamma_{R}$ = 1.01, $\gamma_{I}$ = 1.07. Maguire et al. (2010) also obtained similar results 
(1.02, 0.92, 1.09, respectively). Our estimated $\gamma_{VRI}$ except for SNe 2002hh 
and 2005cs are in good agreements with Patat et al. (1994). Two SNe might have 
additional power sources in this phase (See sections 4.2 for SN 2002hh and 4.5 for SN 2005cs). We find that $\gamma_{JHK}$ are 
$\sim$1-1.4 for all objects. Beyond $\sim$300 days, the fading rates are 
steeper than the $^{56}$Co decay, because the opacity to $\gamma$-rays is decreased and dust starts forming. This 
supports SN dust formation theory. We find flattening or increasing
magnitude features in SNe 1999bw, 2002hh, 2003gd, 2004et, and 2006bc at the same epoch. We will discuss whether these SNe have 
light echos based on the surrounding ISM density later.

\begin{table}
\centering
\caption{The data source of Type {\sc ii} SN and SN 2008S lightcurves. \label{refer}}
\begin{tabular}{@{}ll@{}}
\hline\hline
SN Name   &References\\
\hline
SN 1999bw &Li et al. (2002); Smith et al. (2010); This work\\
SN 2002hh &Pozzo et al. (2006); Welch et al. (2007); This work\\
SN 2003gd &Hendry et al. (2005); Maund \& Smartt (2009); This work\\
SN 2004et &Fabbri et al. (2011a); Kotak et al. (2009); This work\\
SN 2005cs &Pastorello et al. (2009); Tsvetkov et al. (2006); This work\\
SN 2006bc &Brown et al. (2009); Gallagher et al. (2010); This work\\
SN 1987A  &Hamuy \& Suntzeff (1990); Walker \& Suntzeff (1991)\\
SN 1999em &Elmhamdi et al.(2003)\\
\hline
SN 2008S  &Botticella et al. (2009);Prieto et al. (2010); This work\\
\hline
\end{tabular}
\end{table}

\begin{table*}
\centering
\small
\caption{The magnitude decline rate ($\gamma_{band}$) of Type {\sc ii} SNe. \label{dlm}}
\begin{tabular}{@{}l@{\hspace{6pt}}l@{\hspace{6pt}}c@{\hspace{6pt}}c@{\hspace{6pt}}
c@{\hspace{6pt}}c@{\hspace{6pt}}c@{\hspace{6pt}}c|@{\hspace{6pt}}c@{\hspace{6pt}}c@{}}
\hline\hline
$\gamma_{band}$ & Epoch & SN1999bw & SN2002hh & SN2003gd & SN2004et & SN2005cs & SN2006bc & SN1987A & SN1999em \\
\hline
$\gamma_{V}$ & 30-100 & \nodata & \nodata & \nodata & 0.764 & 0.067 & \nodata & \nodata & 0.400 \\
 & 150-300 & \nodata & 0.972 & 0.780 & 1.006 & 0.277 &0.709 & 0.980 & 0.932 \\
 & 300-800 & \nodata & \nodata & \nodata & 1.268 & 1.069 & {\bf 0.333} & 1.469 & 0.985 \\
 & $>$800 & {\bf 0.064} & {\bf 0.008} & \nodata & {\bf 0.173} & {\bf 0.506} & \nodata & 0.562 & \nodata \\
$\gamma_{R}$ & 30-100 & \nodata & 0.147 & \nodata & 0.417 & --0.591 & \nodata & \nodata & --0.030 \\
 & 150-300 & \nodata & 0.971 & 1.211 & 1.014 & 0.684 & \nodata & 0.855 & 0.945 \\
 & 300-800 & \nodata & \nodata & 1.453 & 1.509 & 1.135 & \nodata & 1.640 & 1.126 \\
 & $>$800 & \nodata & \nodata & {\bf 0.052} & \nodata & \nodata & \nodata & 0.528 & \nodata \\
$\gamma_{I}$ & 30-100 & \nodata & \nodata & 0.798 & 0.105 & --0.856 & \nodata & \nodata & --0.269 \\
 & 150-300 & \nodata & 0.968 & 1.295 & 0.918 & 0.771 &0.669 & 0.946 & \nodata \\
 & 300-800 & \nodata & 0.135 & \nodata & 1.454 & {\bf 1.002} & {\bf 0.468} & 1.661 & 1.126 \\
 & $>$800 & \nodata & {\bf 0.049} & {\bf 0.052} & {\bf 0.126} & \nodata & \nodata & 0.565 & \nodata \\
$\gamma_{J}$ & 30-100 & \nodata & \nodata & \nodata & --0.467 & --1.164 & \nodata & \nodata & --0.371 \\
 & 150-300 & \nodata & 1.256 & \nodata & 1.497 & 1.071 & \nodata & 1.192 & 1.158 \\
 & 300-800 & \nodata & \nodata & \nodata & 1.016 & {\bf 1.017} & {\bf 0.493} & 1.528 & \nodata \\
 & $>$800 & \nodata & {\bf 0.007} & \nodata & {\bf 0.117} & \nodata & \nodata & 0.809 & \nodata \\
$\gamma_{H}$ & 30-100 & \nodata & \nodata & \nodata & --0.269 & --0.911 & \nodata & \nodata & --0.428 \\
 & 150-300 & \nodata & 1.316 & \nodata & 1.438 & \nodata & \nodata & 1.305 & 1.444 \\
 & 300-800 & \nodata & \nodata & \nodata & 1.098 & \nodata & {\bf 0.555} & 1.386 & \nodata \\
 & $>$800 & \nodata & {\bf 0.014} & \nodata & {\bf 0.003} & \nodata & \nodata & 0.857 & \nodata \\
$\gamma_{K}$ & 30-100 & \nodata & \nodata & \nodata & --0.245 & --0.784 & \nodata & \nodata & --0.425 \\
 & 150-300 & \nodata & 1.286 & \nodata & \nodata & 1.790 & \nodata & 1.578 & 1.482 \\
 & 300-800 & \nodata & \nodata & \nodata & 1.177 & \nodata & {\bf 0.493} & 1.594 & \nodata \\
 & $>$800 & \nodata & {\bf 0.011} & \nodata & {\bf --0.510} & \nodata & \nodata & 0.624 & \nodata \\
\hline
\end{tabular}
\tablecomments{The $\gamma_{band}$ is the dropping magnitude per a 100 day. 
The values indicated by bold face are derived from the data taken by our observations and the archival $HST$ data.}
\end{table*}

\section{Analysis of late time light curves}

In our prior work on the late time light curves of these  
SNe,  we have concentrated on the dust formation process and 
estimated dust mass as noted in section 4.   
In this paper, we analyze these light curves for two other quantities:  
the  ejected $^{56}$Ni mass (Table \ref{bb}) and the density of the environment when a 
light echo is present (Table \ref{sum}).  
Table \ref{sum} summarizes the derived physical 
parameters for these SNe including  $^{56}$Ni mass, the environment 
density, dust produced and progenitor mass.  Below we describe 
how we calculate the $^{56}$Ni mass and the environment density.

\subsection{Mass of $^{56}$Ni}
Here, we focus on the $^{56}$Ni mass synthesized in the SN explosion, 
because we can estimate this mass using bolometric lightcurves during radioactive decay. 

The method for $^{56}$Ni mass estimation is to (1) determine the bolometric luminosity light curve for the SNe and (2) 
compare the slope during radioactive decay between targets and a standard SN. Here, we use SN 1987A as a standard. 
Arnett \& Fu (1989) estimated 0.073 $\pm$ 0.015 $M_{\odot}$ of freshly synthesized $^{56}$Ni in 
SN 1987A by comparing the observed bolometric luminosity curve and theoretical model.

By using simple blackbody fitting techniques, we estimated the luminosity $L_{\rm hot}$ and emitting radius $R_{\rm hot}$
of the hot component $T_{\rm eff}$ ($>$5000 K here). We estimated these three quantities based on $BVRIJHK$. 
Actually, there are archival $UV$ data only in very early phase. For SN 2004et, Fabbri et al. (2011) 
performed multi components blackbody fittings to the $BVRIJHK$ and {\it Spitzer} IRAC 4 bands and MIPS 24$\mu$m data. 
They showed that the percentages of hot component to the total luminosity exceeded 76 $\%$ from days 64 to 464. 
We assume that the condition is the same as in the other SNe, i.e., the
hot component represented by optical- to near-IR SED is the main component of the bolometric luminosity. 
Since we are interested in the late epoch spectral energy distributions
(SEDs), we selected days 150, 300, and 600. The flux densities at each band are estimated
using the fading rate in Table \ref{dlm}. The results of SED fittings are summarized 
in Table \ref{bb}. The uncertainties of $L_{\rm hot}$ and $T_{\rm eff}$ are $\sim$20$\%$ of the estimated values 
and $\pm$ $\sim$500 K.

Next, we estimated the ejected $^{56}$Ni mass based on the assumption that the energy source
of our sample SNe beyond 150 days is mainly from the power of gamma rays emitted through 
$^{56}$Co to $^{56}$Fe radioactive decay. As the ejecta expand, their opacity to 
$\gamma$-rays is expected to decrease, which results in a modified light curve given by Woosley et al. (1989),
\begin{equation}
\label{co56}
L_{\rm total} = Const \times \exp\left(-\frac{t}{t_{56}}\right)\left[1-\exp\left(-\kappa_{56,\gamma}\phi_{0}
\left(\frac{t_{0}}{t}\right)^{2}\right)\right],
\end{equation}
\noindent  where $t_{56}$ is the $e$-folding life time of $^{56}$Co and the term in the brackets is the effective opacity;
$\kappa_{56,\gamma}$ = 0.033 cm$^{2}$ g$^{-1}$ is the average opacity
to $^{56}$Co-decay $\gamma$-rays, and $\phi_{0}$ = 7$\times$10$^{4}$ g
cm$^{-2}$ is the column depth at the fiducial time $t_{0}$ = 11.6
days. Fitting equation (\ref{co56}) to the estimated bolometric
luminosity curve, we can determine the constant for each SN. 
Assuming that radioactive materials were powering the late time photometric evolution, 
we estimated the $^{56}$Ni mass by comparing the luminosity curves of each SN 
and SN 1987A by the following equation
\begin{equation}
M_{^{56}Ni}(SN) = M_{^{56}Ni}(87A) \times \frac{Const_{SN}}{Const_{87A}}~~~M_{\odot}.
\end{equation}
If we can estimate luminosities at one epoch during the radioactive decay phase, 
we can estimate the $^{56}$Ni mass in principle.  The estimated $^{56}$Ni mass is listed in the last column 
of Table \ref{bb}.  
Since several bands are lacking for SN 2006bc during the radioactive decay phase, 
the estimated $^{56}$Ni mass is based only upon the $HST$ data.
Since the quantities for SN 2002hh are derived from the bolometric luminosity and flux densities of the SN and the light echo, 
the estimated values might have $>$30 $\%$ uncertainties. As later
discussed, the $^{56}$Ni mass of SN 2002hh might go down by $\sim$80
$\%$ (i.e., $\sim$2.8 $M_{\odot}$) when we consider the contribution
from the light echo to the total flux.

\subsection{Light echoes and environment densities}

Light echoes are created when the flash from the SNe scatters off dust in their environments, 
which can be interstellar or circumstellar. The timing of the light echo depends on 
the location of the dust with respect to the SN.  However, light echoes at late times in the SN 
light curve are more easy to detect because the increased light due to the SN is 
more apparent when the SN light has decayed significantly.

The intensity of the light echo depends on the density of the surrounding medium.  
Using the scattering optical depth of the light echo given by Romaniello et al. (2005),
\begin{eqnarray}
\tau_{echo,sca}(\lambda) &=& 7.6 \times 10^{-5}\frac{5.8 \times 10^{21}}{k}
\frac{R(\lambda)}{3.1}\frac{\delta t_{\rm SN}(\lambda)}{\rm 100~~days}\nonumber\\
&&\times\frac{\omega(\lambda)}{0.6}\frac{n({\rm H})}{\rm 1 cm^{-3}}\frac{n({\rm O})}{n({\rm O})_{\odot}},
\end{eqnarray}
where $k$ = $N_{\rm H}$/$E(B-V)$ is the ratio of total neutral hydrogen
column density to color excess at solar metallicity ($k$=5.8$\times$
10$^{21}$ cm$^{-2}$ mag$^{-1}$; Bohlin et al. 1978), $\delta$$t_{\rm SN}$
is the duration of the burst of the SN ($\sim$100 days in the case of
Type {\sc ii}-P), $\omega$($\lambda$) is the grain albedo ($\simeq$0.6 at optical wavelength;
e.g., Mathis et al. 1977). Using equation (2), $\tau_{echo,sca}$($V$)
is estimated to be 1.205$\times$10$^{-4}$$n_{\rm H}$ adopting $\omega(V)$=0.6
and $\delta t_{\rm SN}(\lambda)$=100 days. Sparks (1994) argued that
the difference between peak ($M_{\rm SN}$) and
light echo magnitudes ($M_{\rm echo}$) is
\begin{equation}
M_{\rm echo} \simeq M_{\rm SN} + 0.5 -2.5\log(\tau_{\rm echo,sca}).
\end{equation}
If a  light echo is associated with a SN, $M_{\rm echo}(V)$ -- $M_{\rm SN}(V)$ 
$\simeq$ 10.3 -- 2.5$\log$~$n$(H).  Using the light curves, we can measure 
the magnitude difference in SNe with light echoes and estimate the
density $n({\rm H})$.

\begin{table}
\centering
\caption{The luminosity, temperature, emitting radius, and $^{56}$Ni mass \label{bb}}
\begin{tabular}{@{}lcccccc@{}}
\hline\hline
SN Name      &Epoch&$L_{\rm hot}                   $&$T_{\rm eff}$&$R_{\rm hot}$&$M$($^{56}$Ni)\\
            &(day)&($\times$10$^{6}$ $L_{\odot}$) &(K)
            &($\times$10$^{13}$ cm)&($\times$10$^{-2}$ $M_{\odot}$)\\
\hline
SN2002hh &150  & 238     &5490   &119&14&\\
        &300  & 39.9    &6170   &38.6&\nodata&\\
SN2003gd &300  &2.4&5930&10&0.8\\
SN2004et &150  &92.9   &7040   &45.1&5.9\\
        &300  &12.6     &7200   &15.9&\nodata\\
        &600  &0.3    &6360   &3.26&\nodata\\
SN2005cs   &300  &1.50   &7250   &5.41&0.5\\
SN2006bc   &600  &0.3  &6800   &2.84 &2.7\\
SN1987A    &150  &86.5   &6030   &59.4&7.3\\
        &300  &25.8     &7130   &23.2&\nodata\\
        &600  &0.4    &6500   &3.42&\nodata\\
SN1999em   &150  &22.0   &5560   &35.2&1.9\\
        &300  &6.1     &5220   &21.1&\nodata\\
\hline
\end{tabular}
\tablecomments{The uncertainties of $L_{\rm hot}$ and $T_{\rm eff}$ are $\sim$20$\%$ and $\sim$500 K.}
\end{table}

\section{Results for individual SNe}

\subsection{SN 1999bw}
\begin{figure}
\centering
\includegraphics[bb=29 365 420 692,scale=0.6]{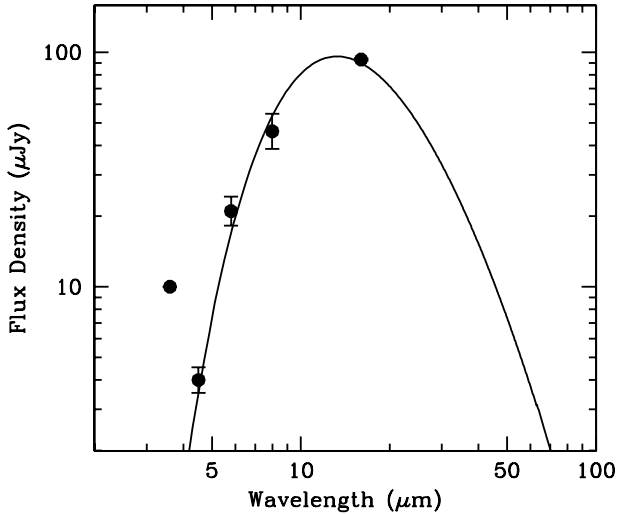}
\caption{The mid-IR SED of SN 1999bw at days 3316 (3.6/4.5/5.8/8.0 $\mu$m) and 3331 (16 $\mu$m). The filled
circles are observations. The thick line is the result of modified blackbody fitting. 
\label{sed99}}
\end{figure}

SN 1999bw is a Type {\sc ii}-n SN in NGC 3198 ($d$=13.7 Mpc;
Freedman et al. 2001) and was discovered on JD.2451288.7.
The position measured from our HST observation 
is in agreement with Sugerman et al. (2004) within 0.5$''$. We detected
the light of the SN in F606W and F814W.
The optical spectra at early epochs are dominated by emission lines having 
full width at half maximum (FWHM) of $\sim$1000-2000 {\kms} (Garnavich et al. 1999; Filippenko
et al. 1999). With {\it Spitzer}/IRAC flux densities and the
blackbody fitting method, Sugerman et al. (2004) estimated the
emitting radius of 1.6$\times$10$^{16}$ cm and the expansion
velocity of 1000 {\kms} in the fifth year after the initial explosion. 
Dust production has not been measured for this SN so far, most likely because 
insufficient data exist for such an estimate. 
Fabbri (2011b) measured flux
densities in 4 Spitzer/IRAC bands (days 3316), IRS-PU 16 $\mu$m (3331),
and MIPS 24 $\mu$m (3320), as listed in
Table \ref{99bw_tab}. Based on her results, we estimated a dust mass of
9.2$\times$10$^{-5}$ $M_{\odot}$ at a temperature of $\sim$300 K
at day $\sim$3320 using modified blackbody fitting, assuming amorphous carbon grains only with 0.1 $\mu$m radius.
In the fitting, we adopted the optical constants of Zubko et al. (1996). 
The best fit SED is presented in Fig.\ref{sed99}.

Smith et al. (2010) reported $BVRI$ magnitudes at the very early plateau phase. 
They suggested that this object might be an impostor from statistical study of 
expansion velocities of Type {\sc ii} SNe and luminous blue variables (LBVs) based 
on optical spectra. They estimated a peak absolute $R$-band magnitude of --12.65 
using optical spectra at day 4, which lies between those of P-Cygni and $\eta$ Carinae.
There is no information on the progenitor star.
There is no information on the brightness of the SN during days 150-626, so 
we estimate the $V$-band magnitude decline rate for $>$800 days.
Since the magnitude difference between peak and day $>$1000 is $\sim$6.6, the hydrogen
density $n$(H) assuming solar O abundance (8.69; Lodders et al. 2003) is estimated to be 
$\sim$30 cm$^{-3}$, which is larger than typical interstellar
gas densities (1-10 cm$^{-3}$). We might therefore be looking mostly at
pre-existing circumstellar material rather than the ISM.

\begin{table}
\centering
\scriptsize
\caption{SN 1999bw $Spitzer$ observations}
\begin{tabular}{@{}cccccccc@{}}
\hline\hline
Obs.Date &Epoch&\multicolumn{6}{c}{Flux density ($\mu$Jy)}\\
         &(day)&3.6 $\mu$m&4.5 $\mu$m&5.8 $\mu$m&8.0 $\mu$m&16 $\mu$m &24 $\mu$m\\
\hline
2008/05/13 &3316&$\le$10 & 4.0$\pm$0.5 & 21$\pm$3 & 46$\pm$8 & $\cdots$ & $\cdots$  \\ 
2008/05/17 &3320&$\cdots$ &$\cdots$      & $\cdots$  &$\cdots$   &$\cdots$  &$\le$650\\
2008/05/28 &3331&$\cdots$ & $\cdots$     &$\cdots$ &$\cdots$&93$\pm$16 &$\cdots$\\
\hline
\end{tabular}
\tablecomments{
The upper limits for the MIPS 24 $\mu$m data were estimated by measuring the flux density of the faint point source seen 
$\sim$6$\arcsec$ to the east of the SN position. This was found to have an average brightness of 657 $\mu$Jy 
with a standard deviation of 2.5 \% 
over the four epochs, and this approximate value was used as a robust upper limit for the SN brightness at each epoch. This table is
taken from Fabbri (2011b). \label{99bw_tab}}
\end{table}

\subsection{SN 2002hh}
\begin{figure}[t]
\centering
\includegraphics[scale=0.55,clip]{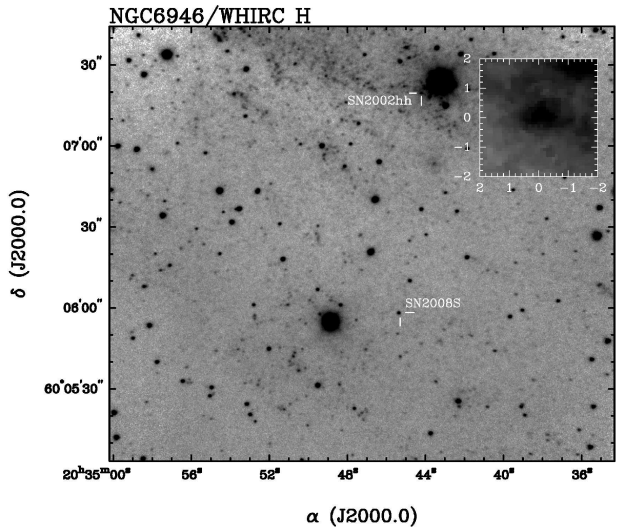}
\caption{WHIRC view of the SNe 2002hh and 2008S
field in the H-band. The positions of SNe 2002hh and 2008S are indicated by the white 
tick marks. A close-up image of SN 2002hh is presented in the inner box
 (2$\arcsec$$\times$2$\arcsec$). The image was taken in a 2008 observing run (PI: M. Meixner).\label{whirc_wfv}}
\end{figure}

SN 2002hh is a Type {\sc ii}-P SN in a spiral arm of the starburst galaxy
NGC 6946 ($d$=5.9 Mpc; Karachentsev et al.  2000) and was discovered
on JD.2452574.5.
We detected the light from the SN in $HST$, NIRI, and WHIRC observations.
The SN birth rate in this galaxy is very high,
$\sim$0.1 SN yr$^{-1}$ during the past  100 yr. Within 2$\arcmin$
radius from SN 2002hh, SNe 1969P and 2008S were observed (Fig.\ref{whirc_wfv}). The high SN
formation rate could be a natural consequence of the high star forming activity in NGC 6946.

\begin{figure*}[t]
\centering
\includegraphics[scale=1.5]{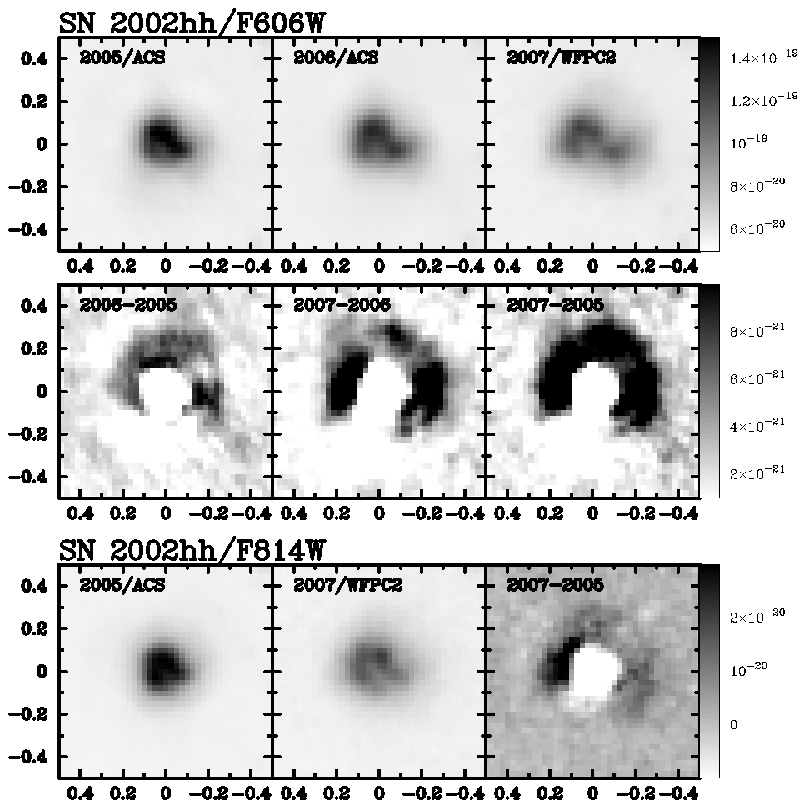}
\caption{({\it upper six panels}) The temporal evolution of SN 2002hh at F606W band. 
({\it lower three panels}): The temporal evolution of SN 2002hh at F814W band. 
The F606W difference images are created after PSF matching each band image. \label{02echo}}
\end{figure*}

SN 2002hh is still bright and close to peak brightness even after $\sim$7 years
(Fig.3). While at the early plateau phase around day 150 the fading rate roughly
followed the $^{56}$Co decay power (0.97 at $V$ and $I$ and $\sim$1.3 at $JHK$),
the fading rates are $\lesssim$0.1 from day 300 or later. Pozzo et al.
(2006) argued that the near-IR excess in this SN is due to an IR light echo from a
pre-existing, dusty circumstellar medium. In Fig.\ref{02echo}, we show the temporal evolution of the SN at F606W ($V$-band).
The 2005 and 2006 images (upper left two panels)
were taken by the ACS/HRC. 
In the middle three panels, we show the subtraction images between two different epochs.
Two arc structures are seen and appear to be expanding with time. We
interpret these structures as light echoes.

Our formula for the density of the environment depends on the metallicity, here measured as the oxygen 
(O) abundance.  Belley \& Roy (1992) investigated
the oxygen (O) abundances in 160 H~{\sc ii} regions in NGC 6946 and estimated the O abundance
gradient of $\Delta$$\log$~($n$(O)/$n$(H))/$\Delta$$R$=--0.089 dex kpc$^{-1}$, where n(O)
and n(H) are the number densities of oxygen and hydrogen and the extrapolated
$\log$~$n$(O)/$n$(H)+12=9.36 and $R$ is the distance from the core of
the galaxy. SNe 2002hh and 2004et (see below) are located in $\sim$5.7 and $\sim$8 kpc
from the center, respectively. When we adopt the solar O abundance,
the O abundances of SNe 2002hh and 2004et are expected to be 8.85 and 8.65, respectively.

Since the difference $V$ magnitude between the peak and day 1000 or later
is $\sim$3.7, $n({\rm H})$ $\sim$440 cm$^{-3}$ is required if the light echo is associated with the SN.
This hydrogen density is close to relatively dense gas density in such
H~{\sc ii} regions rather than that in interstellar space. 
Welch et al. (2007) suggests that most of the $Spitzer$ mid-IR flux may have
come from dust in the star formation region associated with SN 2002hh
precursor. We should note that we measured the magnitude of the SN and
the light echo. In the case of SN 2004et, which has the light echo and 
is a member of NGC 6946, the apparent radius of the SN is $\sim$0.1$\arcsec$. 
If we adopted 0.1$\arcsec$ radius in 
aperture photometry for SN 2002hh and we define that the inner radius of
light echo is $>$0.1$\arcsec$, $\sim$80$\%$ of the total flux 
is from the light echo as of days 1717 (our $HST$ observations). 

 Barlow et al. (2005) estimated 0.1-0.15$ M_{\odot}$ dust around SN 2002hh
after 600 days with radiative transfer modelings based on $Spitzer$/IRAC/MIPS.
However, the minimum emitting radius $\sim$10$^{17}$ cm is
too large for the emitting dust to be forming dust in the SN ejecta.
Pozzo et al. (2006) suggest that dust within $\sim$10$^{17}$ cm
would be evaporated. Barlow et al. (2005) therefore concluded
that the emitting dust must have been pre-existing and
its origin may be from an enhanced mass-loss by the $>$10 $M_{\odot}$ progenitor.

\subsection{SN 2003gd}
SN 2003gd is a Type {\sc ii}-P SN in NGC 628 ($d$=7.2Mpc; Van Dyk et al. 2003)
and was discovered on JD.2452803.2. 
Van Dyk et al. (2003) and Smartt et al. (2004) identified the
progenitor with an F606W $HST$ image; the progenitor is a 6-12 $M_{\odot}$
red supergiant. Maund \& Smartt (2009) estimated the
$V$ and $I$ band magnitudes of the progenitor using pre-explosion
images from the $HST$ and Gemini telescope archives. They estimated
$V$=25.8 and $I$=23.3 and an initial mass of 6-12 $M_{\odot}$.

The $VR$-band fading rates from day 150-300 are comparable with
the radioactive decay rate as earlier mentioned.
But, in days 300-800, the fading rates are much larger than those supported by
the $^{56}$Co decay power, indicating that dust formation had started from
day $\sim$300 or later. Sugerman et al. (2006) estimated the dust mass 
of 1.7$\times$10$^{-3}$ $M_{\odot}$ at day 499 (clumped dust distribution model).

\begin{figure*}[t]
\centering
\includegraphics[scale=0.8]{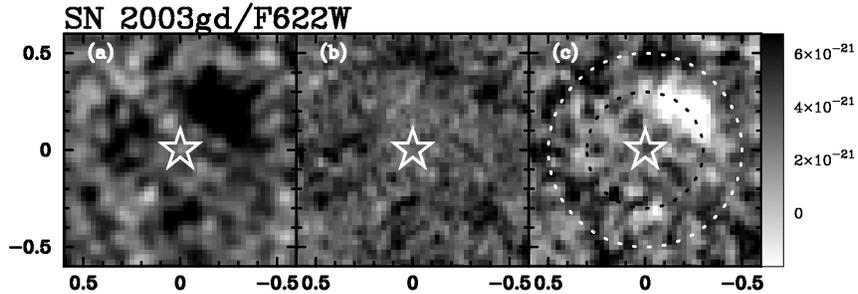}
\caption{The temporal evolution of SN 2003gd light echoes at the F625W band.  
The echo is seen in 2004 ACS (northwestern arc structure, {\it left
 panel}) and 
in 2007 WFPC2 observations (northern 
arc structure, {\it middle panel}). The radius of the echo is $\sim$0.3$''$ in 2004 
and 0.5$''$ in 2007. The position of the SN is indicated by 
the white stars. The difference images in {\it right panel} are created 
after PSF matching each band image. The radii of broken black and white are 
0.3$''$ and 0.5$''$, respectively. 
\label{echo}}
\end{figure*}

When we focus on day $\sim$500 or later,
the magnitude is almost constant. This is due to a light echo.
Sugerman (2005; 0.225$''$ inner and 0.375$''$ outer radii at day 632)
and Van Dyk et al. (2006; 0.31$''$ at the same epoch)
found the light echo in the F625W band. We detected an arc shaped light
echo with  $\sim$0.5$''$ radius (5.39$\times$10$^{19}$ cm at 7.2 Mpc) in the same band
as shown in Fig.\ref{echo}. We should note that the shape of the light echo is changing with time. 
Belley \& Roy (1992) measured the O abundances in 130 H~{\sc ii} regions in NGC 628 and estimated the O abundance
gradient of $\Delta$$\log$~($n({\rm O})/n({\rm H})$)/$\Delta$$R$=--0.081 dex kpc$^{-1}$, where
the extrapolated $\log$~$n({\rm O})/n({\rm H})$+12=9.2 in the galaxy core. When we adopt
$R$ of $\sim$4.5 kpc and $R(R)$=2.32, $M_{\rm echo}(R)$--$M_{\rm SN}(R)$ $\simeq$
10.6-2.5$\log$~$n$(H). Since the difference magnitude between the peak 
and day 1998 is $\sim$10.97, $n$(H) is $\sim$1.4 cm$^{-3}$, which corresponds to a typical
interstellar medium density.

\subsection{SN 2004et}
SN 2004et is a Type {\sc ii}-P SN in NGC 6946  discovered on JD.2453217.5.
We successfully detected the SN at all the observed bands.

The light curves and fading rates at all bands are very similar
to those of SNe 1987A and 1999em except for day $\sim$1000 or later. Up to 1000 days
after the initial explosion, the brightness at both optical and infrared
wavelength bands was monotonically fading, but from $\sim$1200 days it
starts to increase again, in particular in the $K$-band. Our $HST$ observations were taken just 
before and after the beginning of this brightening. Kotak et al. (2009) argued that this phenomenon
is caused by an interaction between the SN ejecta and the surrounding ISM.
On the other hand, Sugerman et al. (in prep.) propose a light-echo as an 
explanation for the re-brightening. Since the difference of $M(V)$ between the plateau and 
day $\sim$1000 is $\sim$10.53, $n$(H) is $\sim$1.2 cm$^{-3}$. 
The magnitude after day $\sim$1000 is consistent with the light echo interpretation.

With the bolometric light-curve Sahu et al. (2006) estimated a $^{56}$Ni mass 
of $\sim$0.06 $M_{\odot}$.  The progenitor mass is 15-24.5 $M_{\odot}$ (Li et al. 2005;
Utrobin \& Chugai 2009). Fabbri et al. (2011a) estimated  an SN produced  dust mass 
of 4$\times$10$^{-4}$ to 4$\times$10$^{-3}$ $M_{\odot}$ using SED fits 
of combined $Spitzer$ and optical/near-IR data. There are active 
star forming regions around SN 2004et. Like SNe 2002hh and 2008S 
(but a SN impostor, see below), the progenitor of 
SN 2004et would have formed in a metal-rich environment.

\begin{table*}[t]
\centering
\small
\caption{Derived quantities for our targets in this study and our SEEDS sample. \label{sum}}
\begin{tabular}{@{}c@{\hspace{5pt}}c@{\hspace{4pt}}c@{\hspace{4pt}}
c@{\hspace{4pt}}c@{\hspace{4pt}}c@{\hspace{4pt}}c@{\hspace{4pt}}c@{\hspace{4pt}}c@{\hspace{4pt}}l@{}}
\hline\hline
SN&Type &Distance&$E(B-V)$&$M_{prog}$   &       $M$($^{56}$Ni) &       $M_{dust}$       &       ISM dens.       &       Light   &       References for\\
Name&   &(Mpc)  &(MW+Host)&($M_{\odot}$)        &($\times$10$^{-2}$ $M_{\odot}$) &       ($M_{\odot}$)   &       (cm$^{-3}$)     &echo?   &$M_{prog}$/$M$($^{56}$Ni)/$M_{dust}$\\
\hline
1999bw  &{\sc ii}-n or&13.7   &0.01   & ?      &       ?       &       $\sim$10$^{-4}$       & $\sim$30              &       No       &(1)       \\
        &impostor\\
2002hh  &{\sc ii}-P   &5.9    &1.97   &$>$10  &       $<$14      &       0.1-0.15 &       $\sim$400  &       Yes     &(1),(2)       \\
2003gd  &{\sc ii}-P   &7.2    &0.18   &6-12   &       0.8  &
 4$\times$10$^{-5}$-1.7$\times$10$^{-3}$ &       $\sim$1.4 &       Yes     &(1),(3),(4)       \\
2004et  &{\sc ii}-P   &5.9    &0.41   &15-24.5        &       5.9     &
 1.5$\times$10$^{-4}$- 1.5$\times$10$^{-3}$ &       $\sim$1.2 &       Yes     &(1),(5),(6),(7)\\
2005cs  &{\sc ii}-P   &7.1    &0.14   &7-12   &       0.5  &$\le$3.1$\times$10$^{-3}$               &$\sim$0.4              &       No      &(1),(8),(9),(10)       \\
2006bc  &{\sc ii}-P or {\sc ii}-L  &16     &0.52   &$\lesssim$12      &       2.7  &       9$\times$10$^{-4}$      &$\sim$36               &       Yes        &(1),(11),(12)\\
2007it  &{\sc ii}-P   &11.7   &0.13   &16-27  &       9       &       $\sim$10$^{-4}$    & $\sim$19              &      Yes     &(13)       \\
2007od  &{\sc ii}-P   &24.5   &0.13   &?      &       0.26    &       4$\times$10$^{-4}$      &$\sim$3.6               &       Yes     &(14)       \\
1987A   &{\sc ii}-P   &50     &0.19   &16-22       &       7.5     &(3-5)$\times$10$^{-4}$-1.3$\times$10$^{-3}$   &$\sim$1.5               &       Yes     &(15)\\
1999em  &{\sc ii}-P   &7.83   &0.10   &12-14  &       1.9     &       $\gtrsim$1$\times$10$^{-4}$         & $\cdots$              &       No      &(16),(17)\\
\hline
2008S  &impostor   &5.9   &0.34$^{a}$   &6-8  &       0.14     & $<$0.02       & ?              &       Yes    &(18),(19)\\
\hline
\end{tabular}
\tablecomments{The extinction by the host galaxy is not included. The
 estimated dust mass is based on mid-IR data. (In SNe 2002hh and 2008S) Most of
 dust is pre-existing material.}
\tablerefs{(1) This work;(2) Barlow et al. (2005);(3) Sugerman et al. (2006);(4) Maund \& Smartt (2009);(5) Fabbri et al. (2011a)
;(6) Li et al. (2005);(7) Utrobin \& Chugai (2009);(8) Li et al. (2006);
(9) Maund et al. (2005);(10) Fabbri (2011b);(11) Gallagher et al. (2011);(12) Smartt et al. (2009);(13) Andrews et al. (2011)
;(14) Andrews et al. (2010);(15) Ercolano et al. (2007);(16) Elmhamdi et al. (2003);(17) Kozasa et al. (2009);(18) Wesson et al. (2010);(19) Botticella et al. (2009)}
 \end{table*}

\begin{figure}
\centering
\includegraphics[scale=0.45,bb=30 164 442 569]{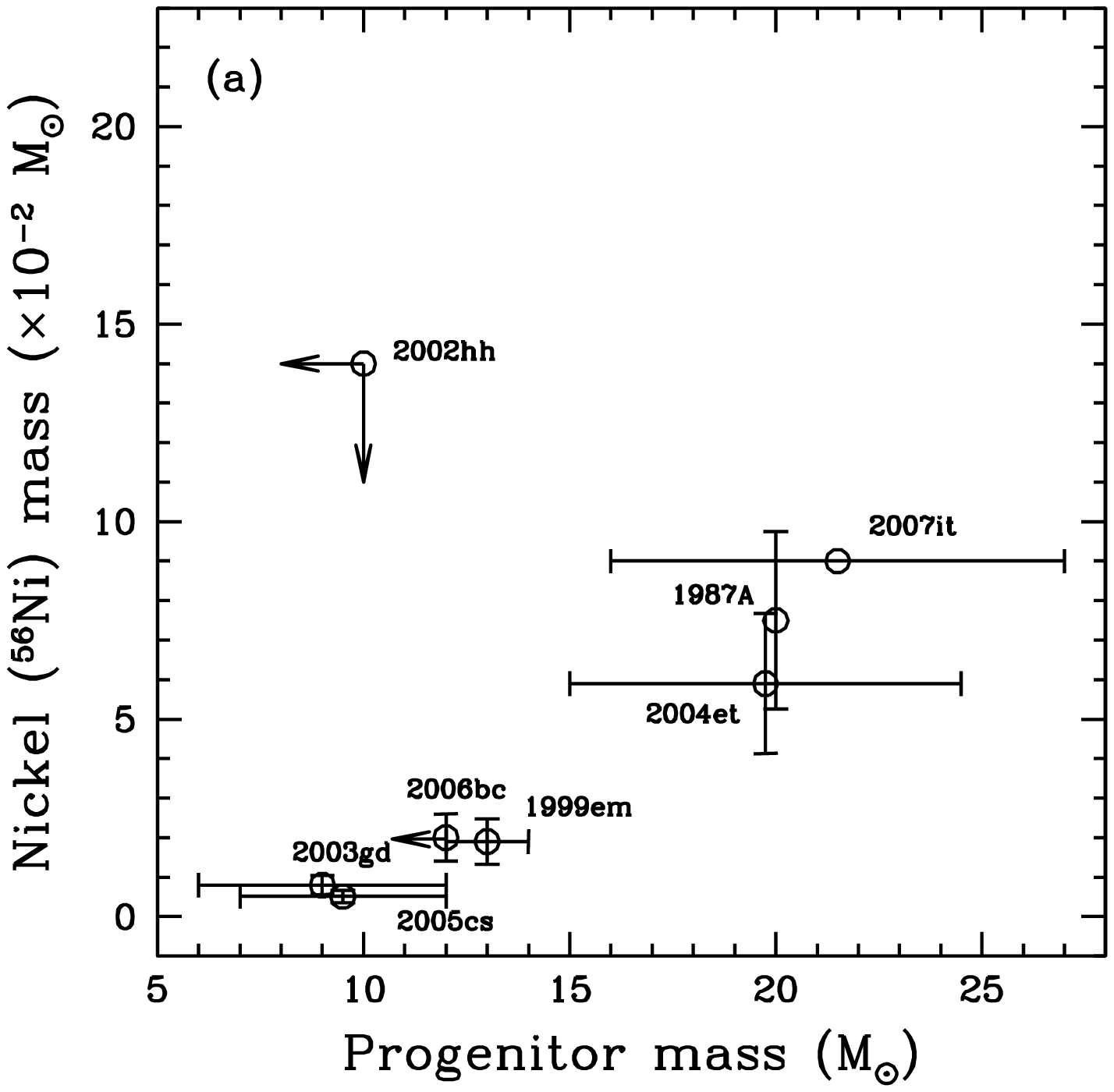}\\
\includegraphics[scale=0.45,bb=30 164 442 569]{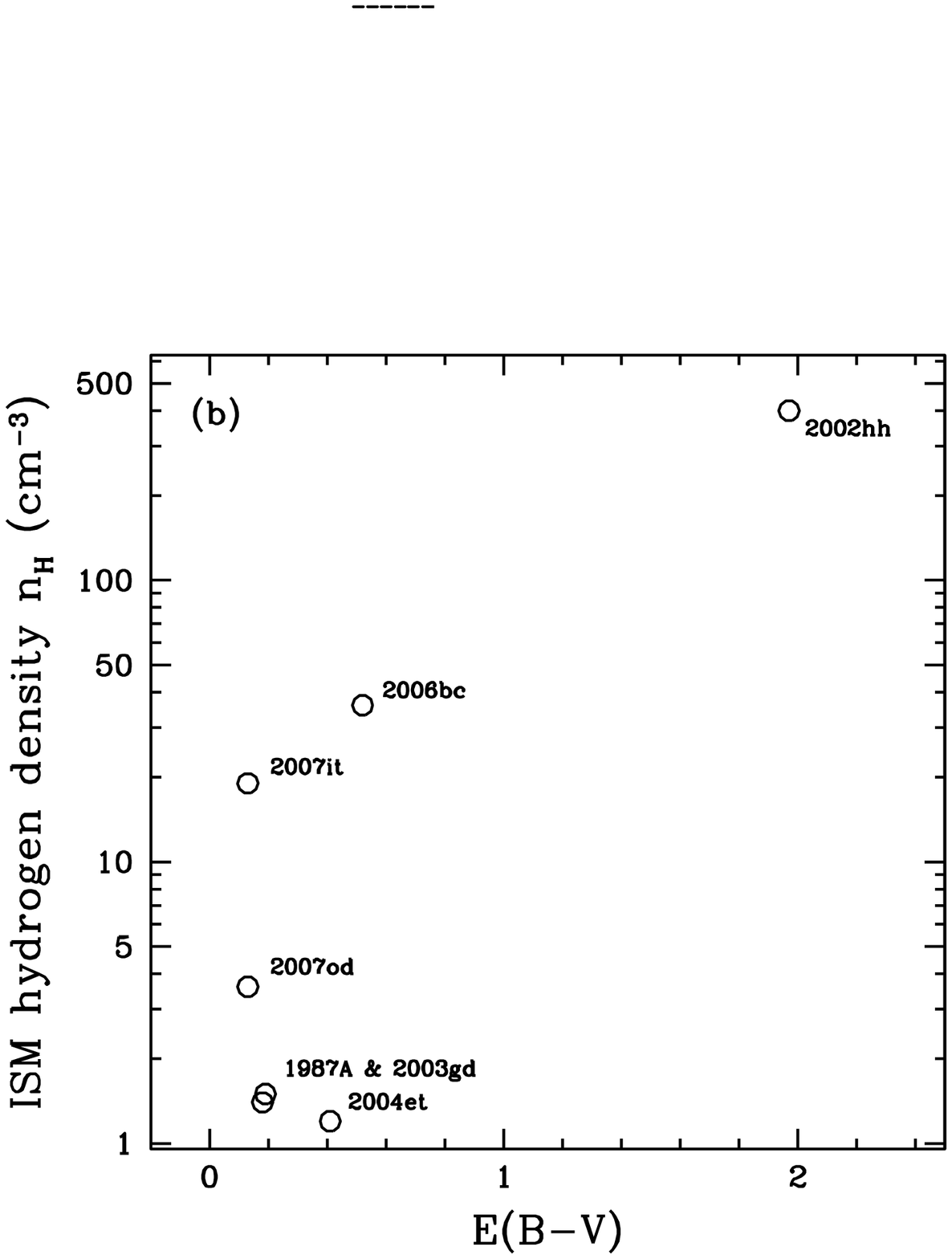}\\
\includegraphics[scale=0.45,bb=30 164 442 569]{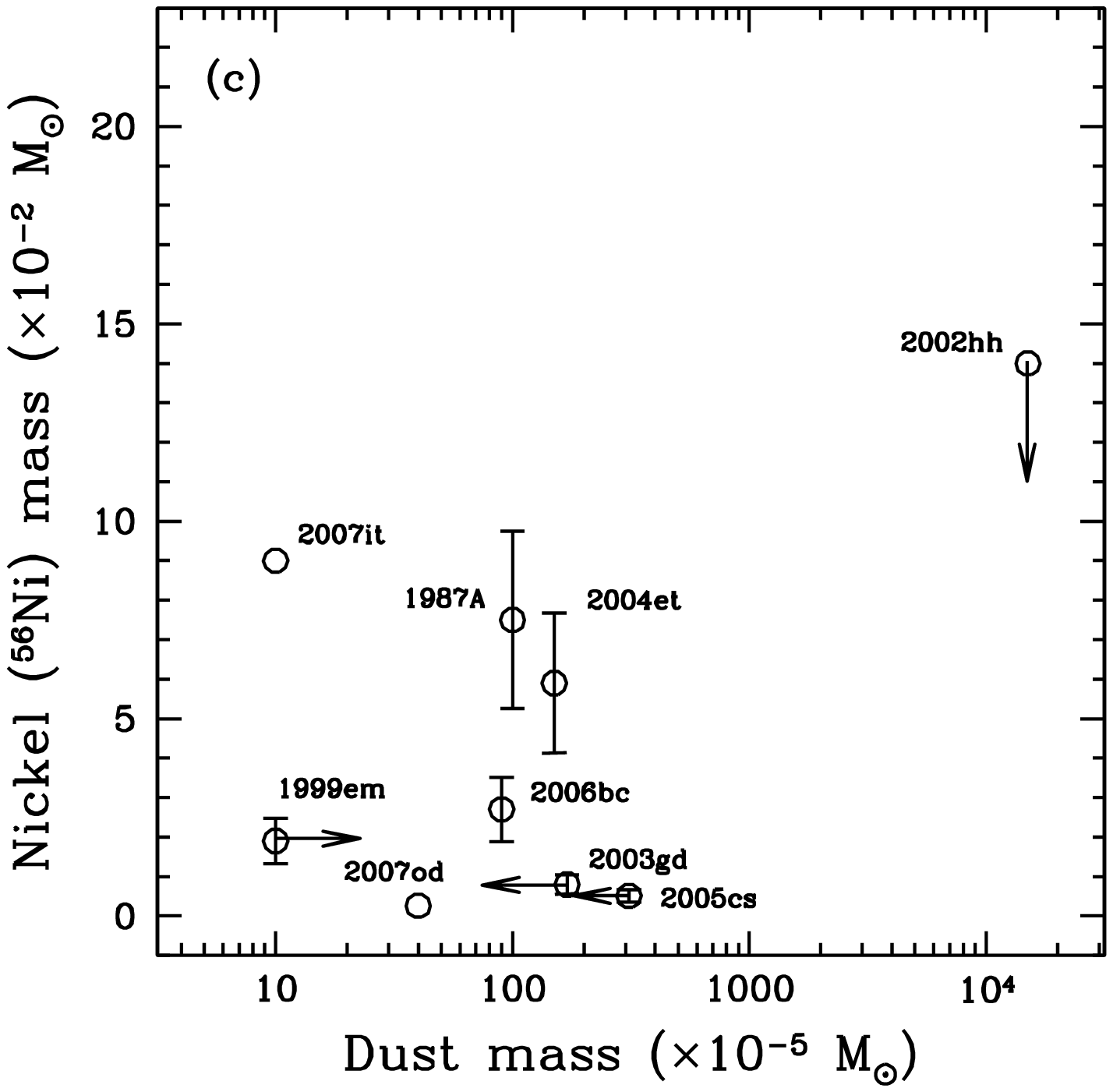}
\caption{({\it upper}) The relationship between nickel $M$($^{56}$Ni) and progenitor masses $M_{prog.}$. ({\it middle}) 
The relationship between ISM hydrogen density and $E(B-V)$. ({\it lower}) The relationship between nickel and dust masses. \label{cor}}
\end{figure}

\subsection{SN 2005cs}
SN 2005cs is a Type {\sc ii}-P SN in NGC 5194 ($d$=7.1 Mpc; Tak{\'a}ts \& Vink{\'o} 2006) and
was discovered on JD.2453550.4. We note that the position measured from the F606W image 
differs significantly from the position we used for observing, i.e. the SIMBAD position. 
As a consequence,  SN 2005cs was unfortunately on the edge of the field for the observations 
we report and was entirely missed by our NIC2 measurement in 2008.
SN 2005cs is $\sim$0.1$''$ northeast from the bright cluster (Li et al. 2006).
Due to its faintness, we can not detect the SN in the NIC2/F160W and 205W bands.

The $VRI$ fading rate at day 150-300 is much smaller than the
$^{56}$Co decay power. Pastorello et al. (2009) argued that such flattening could be
formed by a radiation flow generated in the warmer inner ejecta, which propagates throughout
the transparent cooler external layers, and contributes to the light curve as an additional
source. The lightcurves and magnitude decline rates beyond day $\sim$300 are 
comparable to SNe 2004et and 1987A, suggesting that dust formation may have occurred. 
Fabbri (2011b) gave an upper limit to the dust mass = 3.1$\times$10$^{-3}$ $M_{\odot}$ (O-rich case) 
from the detection limit of a Gemini/Michelle 10 $\mu$m image obtained in 2007 (PI: M.J.Barlow).

Li et al. (2006) detected the progenitor from the $HST$ ACS images, and they estimated $V$$>$25.5 and
$I$$>$24.14. Maund et al. (2005) and Li et al. (2006) reported the detection of the progenitor using
the archival $HST$ ACS images. They argued that the progenitor is a $\sim$7-12 $M_{\odot}$
RSG. Pastorello et al. (2009) obtained  similar results (8-15 $M_{\odot}$).
Pastorello et al. (2009) also estimated that the explosion energy is $\sim$3$\times$10$^{50}$
erg (cf. $\sim$10$^{51}$ erg in typical Type {\sc ii} SNe) and a $^{56}$Ni mass of 
2.8-4.2 $\times$ 10$^{-3}$ $M_{\odot}$. They measured a low expansion velocity (1000-2000
km s$^{-1}$) in optical and near-IR spectra.  The flattening of the  light curve in the last two epoch 
measurements suggests a light echo. Since the difference between the peak and day 
1006 is 11.2 in the $V$-band, the ISM density is $\sim$0.4 cm$^{-3}$ assuming that the SN has a light echo and 
the SN light after day 1006 is partially from the light echo.

\subsection{SN 2006bc}
SN 2006bc is a Type {\sc ii}-L (Gallagher et al. 2010) or Type {\sc ii}-P (Brown et
al. 2009) SN in NGC 2397 ($d$=16 Mpc; Mihos \& Bothun 1997)  and
was discovered on JD.2453819.4. The {\it SWIFT} telescope detected the SN light at an early epoch and 
the $V$-magnitude was $\sim$17.5 at day 15.8 (Brown et al. 2009).
We detected the SN light in all bands. Smartt et al. (2009) give an upper limit of the progenitor 
mass of 12 $M_{\odot}$for SN 2006bc 

SN 2006bc is surrounded by an H\,{\sc ii} region; the optical spectra
taken by Gemini-N/GMOS show the complex of the SN and H\,{\sc ii} region
components (Gallagher et al. 2011). Gallagher et al. (2011) extracted
the SN component by multi-Gaussian fitting, and they confirmed that the 
H$\alpha$ line profile is blueward shifting as the SN becomes older suggesting 
dust production. An IR-excess has been measured in the {\it Spitzer} IRAC bands 
and SED fitting by Gallagher et al. (2011) indicates a 
dust mass of $9\times 10^{-4}$ $M_\odot$. The fading rates 
are much smaller than the typical $^{56}$Co decay power and our estimated $^{56}$Ni 
mass of 2.7$\times$10$^{-2}$ $M_{\odot}$ is lower than average in our sample. 
A flattening of the light curve between days 400 and 500 suggests 
a short duration light echo.  The estimated density of the matter 
in this light echo is $\sim$36 cm$^{-3}$ from the difference
$V$-magnitude = 6.4 between day 15.8 and 694.

\subsection{SN Impostor: SN 2008S}
SN 2008S is regarded as an SN impostor in NGC 6946 and was discovered around 
JD.2454498 (Arbour \& Boles 2008). $E(B-V)$ by the Galaxy is
estimated to be 0.34 from reddening maps of Schlegel et al. (1998). 
Botticella et al. (2009) estimated the extinction by the host galaxy $A_{V}$$\sim$1 (1.13 by the Galaxy) 
by comparing their light echo model with the observed SED at 17 days. 
In this paper, we corrected only the extinction by the Galaxy with $E(B-V)$=0.34.
Our $JHK$ light curves show that SN 2008S is quite different from Type {\sc ii} SNe: for example,
no plateau phase and no clear level-off around day $\sim$100. The nature
of the progenitor star is under debating. Botticella et al. (2009)
argue that SN 2008S may have evolved from a 6-8 $M_{\odot}$ asymptotic giant branch (AGB) star and formed not 
by Fe-core collapse but O-Ne-Mg core collapse by electron capture during the
AGB phase. While, Smith et al. (2010) argue for a high mass progenitor
($>$20 $M_{\odot}$) for this class of objects such as SN2008S.
The $^{56}$Ni mass of SN 2008S has the smallest Ni mass 
estimated for Type {\sc ii} SNe, the value is 1.4 $\times$10$^{-3}$
$M_{\odot}$ (Botticella et al. 2009).

Wesson et al. (2010) estimated the dust mass-loss rate of the progenitor of
SN 2008S to be 5$\times$10$^{-7}$ $M_{\odot}$ yr$^{-1}$ and a total
dust injection into the ISM to be $<$0.01 $M_{\odot}$. 
The large infrared excess on day 17 could 
only be an echo, as there would be no way for dust in the necessary 
quantity to have formed in such a short time. 
The SED on day 180 could be well matched by models using 
the same shell geometry as day 17, without any need for new dust formation.

\section{Discussion}
In Table \ref{sum}, we summarize the derived physical quantities of our sample. 
$^{56}$Ni mass (sixth column of this table) and ISM density (eighth column) 
are from this work. The other quantities are from the literature listed in the last column.

The behaviors of the SN lightcurves are affected 
mainly by (1) the explosion energy, (2) the ISM density ($n_{\rm H}$), and (3) the formed dust mass ($M_{dust}$) in the SNe, 
because the explosion energy determines the peak luminosity 
of the SNe, the surrounding ISM density/distribution and the dust affects the lightcurves at $\gtrsim$1000 days and at $\gtrsim$300 days, 
respectively. Relations between progenitor mass ($M_{prog}$) and $^{56}$Ni mass and explosion energy are known for massive ($>$20 $M_{\odot}$) 
Type {\sc ii} SNe (e.g., Maeda et al. 2010), however, relations among the $^{56}$Ni mass, $M_{dust}$, $M_{prog}$, and ISM density are unknown for 
less massive Type {\sc ii} SNe. Through this work, we have estimated the ISM density and the $^{56}$Ni mass for our sample 
using our $HST$ data. Using the theoretical radiative transfer code MOCASSIN (Ercolano et al. 2005), we have estimated 
dust masses for our sample except for SNe 1999bw, 1999em, and 2005cs. The dust masses listed in this table are based on mid-IR data and 
the value is as of $<$800 days. Here, we will examine relations among $^{16}$Ni mass, $M_{dust}$, $M_{prog}$, ISM density, 
and $E(B-V)$.

\subsection{$^{56}$Ni mass vs. $M_{prog}$} 
In Fig.\ref{cor}(a), we present the relation between the $^{56}$Ni and the progenitor masses $M_{prog.}$. 
In principal, there is a relationship among the amount of $^{56}$Ni yield in stars, the progenitors masses, 
the initial abundances, and the metallicity. The metallicity of our sample is similar, we assume that 
the effects of metallicity in nucleosynthesis would be therefore negligible. 
Maeda et al. (2010) and his colleagues find relationships between the predicted $^{56}$Ni mass (of order 10$^{-2}$ M$_{\odot}$), 
the SN progenitor masses, and the kinematic energy of the explosion (1$\sim$100 $\times$10$^{51}$ erg s$^{-1}$), 
however they focused on Type {\sc ii} SNe evolved from $>$20-50 $M_{\odot}$ progenitors (hypernova). We expected a similar 
relationship between the $^{56}$Ni and progenitor masses in Type {\sc
ii} SNe with progenitor masses of $\sim$10-25 $M_{\odot}$. 
Fig.\ref{cor}(a) strongly supports the previous works. As early mentioned, the estimated $^{56}$Ni mass might have large uncertainties. When we exclude the data of SN 2002hh (due to the large uncertainty of the $^{56}$Ni mass), $M$($^{56}$Ni) 
represents 0.31$\times$10$^{-2}$ $\times$ $M_{prog.}$($M_{\odot}$). The $^{56}$Ni mass derived from 
the bolometric luminosity could be an effective tool to estimate the SN progenitor mass.

 \subsection{$n_{\rm H}$ vs. $E(B-V)$} 
Fig.\ref{cor}(b) shows the relation between ISM density $n_{\rm H}$ and the extinction $E(B-V)$. 
We added information on SNe 2007it (Andrews et al. 2010) and 2007od (Andrews et al. 2011), 
which are targets in our SEEDS project. 
Andrews et al. estimated the nickel mass and dust mass and found the 
light echo by $HST$ observations. The ISM densities around SNe 2007it and 2007od are derived by the same way 
applied to the other SNe, using the difference $V$-band magnitude between plateau and day 922 in 2007od 
(7.12) and day 811 in 2007it (8.9). For both objects, we assume solar metallicity.

There is no correlation between them. This can be explained by the different light echo geometry (shape, size, and 
inclination angle, etc.). For example, modeling of the possible light echo geometries in SN 2004et currently 
suggests an hourglass-shaped nebula similar in size to that discovered around SN 1987A, but oriented such that 
we are looking down the symmetry axis (Sugerman et al. in prep.).

\subsection{$^{56}$Ni mass vs. $M_{dust}$} 
Fig.\ref{cor}(c) shows the relation between the $^{56}$Ni mass and the dust mass $M_{dust}$. The dust mass in this 
diagram is the maximum value in an early phase within 1-3yr. If the correlation between the $^{56}$Ni 
and progenitor masses is true and we exclude SN 2002hh, where we observed the pre-existing 
dust rather than SN origin dust,  there is no relation or a very weak relation 
between the dust and progenitor masses within $\sim$30 $M_{\odot}$. This diagram also implies that 
Type {\sc ii} origin dust mass ranges from $\sim$10$^{-4}$ upto 10$^{-2}$ $M_{\odot}$ per SN at day $<$800 despite the progenitor mass.

The origin of dust in high-$z$ galaxies is a hotly-debated issue stemming from both  
observational and theoretical studies since Bertoldi et al. (2003) discovered 
large amounts of dust ($\sim$4$\times$10$^{8}$ $M_{\odot}$) in the QSO J1148+5251 ({\it z}=6.4, $\sim$900 Myr). 
At the present epoch in our Galaxy, Type {\sc ii} SNe and AGB stars are the main dust producers (Gehrz et al. 1989). 
In young galaxies, AGB stars are not likely to contribute significantly to dust production. Such low-mass 
stars proceed too slowly toward the AGB phase to produce dust within 1 Gyr. For example, 
$\sim$1 $M_{\odot}$ single stars with solar metallicity take $\sim$10 Gyr to evolve into 
the thermally pulsing AGB (TP AGB). If the initial mass is $\sim$5 $M_{\odot}$, these stars can evolve into TP AGB stars and might be able to be the main dust producers 
in young galaxies within $\sim$1 Gyr (Valiante et al. 2009). However, massive stars can evolve into Type {\sc ii} SNe 
in timespans under 20 Myr. Theoretical ISM dust models by Dwek \& Cherchneff (2010) 
predict that an average 20 $M_{\odot}$ (initial mass) SN has to make at least 
$\sim$0.1-1 $M_{\odot}$ of dust in order to be a viable source for the dust 
found in the QSO J1148+5251. However, the $^{56}$Ni mass-$M_{dust}$ diagram does not follow the prediction by 
Dwek \& Cherchneff (2010).

\section{Summary}
We performed $BVRIJHK$ band photometry of 6 Type {\sc ii} SNe at late epochs (days $>$500) using $HST$ 
and ground based telescopes to investigate their natures and relations between dust mass, $^{56}$Ni, ISM density, and 
progenitor mass. Most of our $HST$ observations are successful in detecting the light from
the SNe alone and in measuring magnitudes with less contamination from nearby stars. 
Combining our data with previously published data, we showed lightcurves at $VRIJHK$-bands 
and estimate the decline magnitude rates at each band at 4 different phases. These lightcurves and 
other data indicate that dust is forming in our targets from day $\sim$300-400, supporting
SN dust formation theory.  We estimated the ISM or circumstellar density around SNe 2002hh, 2003gd, 
2004et, and 2006bc. The light echo density around SN 2002hh is higher than a typical 
ISM density. The ISM density around the other SNe is a 1-70 cm$^{-3}$.  We estimated $^{56}$Ni
masses (0.5-14 $\times$10$^{-2}$ $M_{\odot}$) by comparing their bolometric luminosity with SN 1987A 
and we find that it correlates with progenitor mass. This relation supports the work by 
Maeda et al. (2010), who focus on Type {\sc ii} hyper SNe ($>$20 $M_{\odot}$). The dust mass does 
not appear to be correlated with $^{56}$Ni mass among the 7 SNe.

\acknowledgements
M.O. and M.M. acknowledge funding support from STScI GO-1129.01-A and NASA NAO-50-12595. M.O.
acknowledges funding support form STScI DDRF D0101.90128. M.M. appreciates support from Harvard-Smithsonian
Center for Astrophysics during this work. This work is in part based on one of the
author's (JF) dissertation submitted to University College London, in partial fulfillment of the requirement
for the doctorate. This work is in part based on $HST$ archive data
downloaded from the Canadian Astronomy Data Centre.  We thank the support of David Riebel, Lynn Carlson, 
Brian Ferguson, and NOAO operators  for WHIRC observations of these SNe.

%======================================================================
%References
%======================================================================

\end{document}